\documentclass[journal,onecolumn,12pt]{IEEEtran}
\IEEEoverridecommandlockouts
\usepackage{mathpazo}
\usepackage{times}
\usepackage[latin1]{inputenc}
\usepackage[T1]{fontenc}
\usepackage{dsfont, mathrsfs}
\usepackage{fancyvrb, relsize}
\usepackage{amsmath,amssymb,amsfonts}
\usepackage[noend]{algorithmic}
\usepackage{setspace}
\usepackage{algorithm}
\usepackage{subfigure}                                                         
\usepackage{multicol}
\usepackage{graphicx}	
\usepackage{epsfig}
\usepackage{upref}
\usepackage{theorem}
\usepackage{tikz}
\usepackage{stfloats}
\usepackage{enumerate}
\usepackage{cite}
\usetikzlibrary{arrows}
\usetikzlibrary{positioning}
\usepackage[version=3]{mhchem}

\theoremstyle{plain} \theorembodyfont{\normalfont\slshape}
\newcommand{\mathset}[1]{\left\{#1\right\}}

\newtheorem{thm}{Theorem$\!$}
\newenvironment{theorem}
{\begin{thm}\hspace*{-1ex}{\bf.}}{\end{thm}}

\newtheorem{lem}[thm]{Lemma$\!$}
\newenvironment{lemma}{\begin{lem}\hspace*{-1ex}{\bf.}}{\end{lem}}

\newtheorem{prop}[thm]{Proposition$\!$}
\newenvironment{proposition}{\begin{prop}\hspace*{-1ex}{\bf.}}{\end{prop}}

\newtheorem{cor}[thm]{Corollary$\!$}

\newtheorem{defn}[thm]{Definition$\!$}
\newenvironment{definition}{\begin{defn}\hspace*{-1ex}{\bf.}}{\end{defn}}

\newtheorem{xmpl}[thm]{Example$\!$}
\newenvironment{example}{\begin{xmpl}\hspace*{-1ex}{\bf.}}{\end{xmpl}}

\newtheorem{cnstr}[thm]{Construction$\!$}
\newenvironment{construction}{\begin{cnstr}\hspace*{-1ex}{\bf.}}{\end{cnstr}}

\newtheorem{pro}[thm]{Property$\!$}
\newenvironment{property}{\begin{pro}\hspace*{-1ex}{\bf.}}{\end{pro}}

\DeclareMathOperator*{\argmax}{arg\,max}

\begin{document}
\title{Asymmetric Error Correction\\ and Flash-Memory Rewriting  using Polar Codes}
\author{\IEEEauthorblockN{{Eyal En Gad}, {Yue Li}, {Joerg Kliewer}, {Michael Langberg},\\  {Anxiao (Andrew) Jiang} and {Jehoshua Bruck}}
\thanks{The material in this paper was presented in part at the IEEE Int. Symp. on Inform. Theory (ISIT), Honolulu, HI, USA, July 2014~\cite{EngLiKliLanJiaBru14a}. This work was supported in part by Intellectual Ventures, NSF grants CIF-1218005, CCF-1439465, CCF-1440001 and CCF-1320785, NSF CAREER Award CCF-0747415 and the US-Israel Binational Science Foundation (BSF) under Grant No. 2010075.

Eyal En Gad, Yue Li and Jehoshua Bruck are with the California Institute of Technology, Pasadena, CA 91125, \{eengad, yli, bruck\}@caltech.edu.
Joerg Kliewer is with New Jersey Institute of Technology, Newark, NJ 07102, jkliewer@njit.edu.
Michael Langberg is with SUNY at Buffalo, Buffalo, NY  14260 and the Open University of Israel, Raanana 43107, Israel, mikel@buffalo.edu.
Anxiao (Andrew) Jiang is with Texas A\&M University, College Station, TX 77840, ajiang@cse.tamu.edu.}}
\maketitle

\begin{abstract}
We propose efficient coding schemes for two communication settings: 1. asymmetric channels, and 2. channels with an informed encoder. These settings are important in non-volatile memories, as well as optical and broadcast communication. The schemes are based on non-linear polar codes, and they build on and improve recent work on these settings. In asymmetric channels, we tackle the exponential storage requirement of previously known schemes, that resulted from the use of large Boolean functions. We propose an improved scheme, that achieves the capacity of asymmetric channels with polynomial computational complexity and storage requirement. 

The proposed non-linear scheme is then generalized to the setting of channel coding with an informed encoder, using a multicoding technique. We consider specific instances of the scheme for flash memories, that incorporate error-correction capabilities together with rewriting. Since the considered codes are non-linear, they eliminate the requirement of previously known schemes (called polar write-once-memory codes) for shared randomness between the encoder and the decoder. Finally, we mention that the multicoding scheme is also useful for broadcast communication in Marton's region, improving upon previous schemes for this setting.
\end{abstract}

\IEEEpeerreviewmaketitle
\allowdisplaybreaks

\section{Introduction}

In this paper we make several contributions to the design and analysis of error-correcting codes in two important communication settings: 1. asymmetric channel coding, and 2. channel coding with an informed encoder. Asymmetric channel coding is important for applications such as non-volatile memories, in which the electrical mechanisms are dominantly asymmetric~\cite{CasSchBohBru10}. Another important application is in optical communication, where photons may fail to be detected ($1\to 0$) but a false detection when no photon was sent ($0\to 1$) is much less likely~\cite[Section IX]{Gor62}. Channel coding with an informed encoder is also important for non-volatile memories, since the memory state in these devices affects the fate of writing attempts. Channel coding with an informed encoder is also useful in broadcast communication, where it is used in Marton's coding scheme to achieves high communication rates (see~\cite[p. 210]{ElgKim12}).

The focus of this paper is on polar coding techniques, as they are both highly efficient in terms of communication rate and computational complexity, and are relatively easy to analyze and understand. Polar codes were introduced by Arikan in~\cite{Ari09}, achieving the symmetric capacity of binary-input memoryless channels. The first task that we consider in this paper is that of point-to-point communication over asymmetric channels. Several polar coding schemes for asymmetric channels were proposed recently, including a pre-mapping using Gallager's scheme~\cite[p. 208]{Gal68} and a concatenation of two polar codes~\cite{SutRenDupRen12}. A more direct approach was proposed in~\cite{HonYam13}, which we consider in this paper. A similar approach is also considered in~\cite{MonHasUrb14}. The scheme in~\cite{HonYam13} achieves the capacity of asymmetric channels using non-linear polar codes, but it uses large Boolean functions that require storage space that is exponential in the block length. We propose a modification for this scheme, that removes the requirement for the Boolean functions, and thus reduces the storage requirement of the encoding and decoding tasks to a linear function of the block length. 

The second contribution of this paper is a generalization of the non-linear polar-coding scheme to the availability of channel side information at the encoder. We call this scheme a polar multicoding scheme, and we prove that it achieves the capacity of channels with informed encoders. The capacity of such channels was characterized by Gelfand and Pinsker in~\cite{GelPin80}. This scheme is useful for non-volatile memories such as flash memories and phase change memories, and for broadcast channels. We focus mainly on the flash memory application. 

A prominent characteristic of flash memories is that the response of the memory cells to a writing attempt is affected by the previous content of the memory. This complicates the design of error correcting schemes, and thus motivates flash systems to ``erase'' the content of the cells before writing, and by that to eliminate its effect. However, the erase operation in flash memories is expensive, and therefore a simple coding scheme that does not require erasures could improve the performance of solid-state drives significantly. We show two instances of the proposed polar multicoding scheme that aim to achieve this goal.

\subsection{Relation to Previous Work}

The study of channel coding with an informed encoder was initiated by Kusnetsov and Tsybakov~\cite{KusTsy74}, with the channel capacity derived by Gelfand and Pinsker~\cite{GelPin80}. The informed encoding technique of Gelfand and Pinsker was used earlier by Marton to establish an inner bound for the capacity region of broadcast channels~\cite{Mar79}. Low-complexity capacity-achieving codes were first proposed for continuous channels, using lattice codes~\cite{ZamShaEre02}. In discrete channels, the first low-complexity capacity-achieving scheme was proposed using polar codes, for the symmetric special case of information embedding~\cite[Section VIII.B]{KorUrb10}. A modification of this scheme for the application of flash memory rewriting was proposed in~\cite{BurStr13}, considering a model called write-once memory. An additional scheme for the application of flash memory, based on randomness extractors, was also proposed recently~\cite{GabSha12}.

Our work is concerned with a setup that is similar to those considered in~\cite{BurStr13,GabSha12}. An important contribution of the current paper compared to~\cite{BurStr13,GabSha12} is that our scheme achieves the capacity of a rewriting model that also includes noise, while the schemes in~\cite{BurStr13,GabSha12} address only the noiseless case. Indeed, error correction is a crucial capability in flash memory systems. Our low-complexity achievability of the noisy capacity is done using a multicoding technique. Comparing with~\cite{GabSha12}, the current paper allows an input cost constraint, which is important in rewriting models for maximizing the sum of the code rates over multiple rewriting rounds. Comparing with~\cite{BurStr13}, the current paper also improves by removing the requirement for shared randomness between the encoder and the decoder, which limits the practical coding performance. The removal of the shared randomness is done by the use of non-linear polar codes. An additional coding scheme was proposed during the writing of this paper, which also does not require shared randomness~\cite{Ma14}. However, the scheme in~\cite{Ma14} considers only the noiseless case, and it is in fact a special case of the scheme in the current paper.

Polar coding for channels with informed encoders was implicitly studied recently in the context of broadcast channels, as the Marton coding scheme for broadcast communication contains an informed encoding instance as an ingredient. In fact, a multicoding technique similar to the one presented in this paper was recently presented for broadcast channels, in~\cite{GoeAbbGas13}. While we were unaware of the result of~\cite{GoeAbbGas13} and developed the scheme independently, this paper also has three contributions that were not shown in~\cite{GoeAbbGas13}. First, by using the modified scheme of non-linear polar codes, we reduce the storage requirement from an exponential function in the block length to a linear function. Secondly, we connect the scheme to the application of data storage and flash memory rewriting, that was not considered in the previous work. And thirdly, the analysis in~\cite{GoeAbbGas13} holds only for channels whose capacity-achieving distribution forms a certain degraded structure. In this paper we consider a specific noisy rewriting model, whose capacity-achieving distribution forms the required degraded structure, and by that we show that the scheme achieves the capacity of the considered flash-memory model. 

Another recent paper on polar coding for broadcast channels was published recently by Mondelli et. al.~\cite{MonHasSasUrb14}. That paper proposed a method, called ``chaining'', that allows to bypass the degraded structure requirement. In this paper we connect the chaining method to the flash-memory rewriting application and to our new non-linear polar coding scheme, and apply it to our proposed multicoding scheme. This allows for a linear storage requirement, together with the achievability of the informed encoder capacity and Marton's inner bound, eliminating the degraded structure requirement. Finally, we show an important instance of the chaining scheme for a specific flash-memory model, and explain the applicability of this instance in flash-memory systems.
 
The rest of the paper is organized as follows. Section~\ref{sec:asymmetric} proposes a new non-linear polar coding scheme for asymmetric channels, which does not require an exponential storage of Boolean functions. Section~\ref{sec:multicoding} proposes a new polar multicoding scheme for channels with informed encoders, including two special cases for the rewriting of flash memories. Finally, Section~\ref{sec:discussion} summarizes the paper.


\section{Asymmetric Point-to-Point Channels}
\label{sec:asymmetric}

\emph{Notation:} For positive integers $m\le n$, let $[m:n]$ denote the set $\mathset{m,m+1,\dots,n}$, and let $[n]$ denote the set $[1:n]$. Given a subset $\mathcal{A}$ of $[n]$, let $\mathcal{A}^c$ denote the complement of $\mathcal{A}$ with respect to $[n]$, where $n$ is clear from the context. Let $x_{[n]}$ denote a vector of length $n$, and let $x_{\mathcal{A}}$ denote a vector of length $|\mathcal{A}|$ obtained from  $x_{[n]}$ by deleting the elements with indices in $\mathcal{A}^c$. 

Throughout this section we consider only channels with binary input alphabets, since the literature on polar codes with non-binary codeword symbols is relatively immature. However, the results of this section can be extended to non-binary alphabets without much difficulty using the methods described in~\cite{MorTan10,ParBar13,SahPra11,SasTelAri09,Sas12,Sas12a}. The main idea of polar coding is to take advantage of the polarization effect of the Hadamard transform on the entropies of random vectors. Consider a binary-input memoryless channel model with an input random variable (RV) $X\in\mathset{0,1}$, an output RV $Y\in\mathcal{Y}$ and a pair of conditional probability mass functions (pmfs) $p_{Y|X}(y|0),p_{Y|X}(y|1)$ on $\mathcal{Y}$.  Let $n$ be a power of 2 that denotes the number of channel uses, also referred to as the block length. The channel capacity is the tightest upper bound on the code rate in which the probability of decoding error can be made as small as desirable for large enough block length. The channel capacity is given by the mutual information of $X$ and $Y$.
\begin{theorem}
\label{th:capacity}
\textbf{(Channel Coding Theorem)}\cite[Chapter 7]{CovTho06} The capacity of a discrete memoryless channel defined by $p_{X|Y}$ is 
$$C=\max_{p_X}I(X;Y).$$
\end{theorem}

 The Hadamard transform is a multiplication of the random vector $X_{[n]}$ over the field of cardinality 2 with the matrix $G_n=G^{\otimes\log_2 n}$, where $G=\left( \begin{array}{ccc}
1 & 0\\
1 & 1\end{array}\right)$ and $\otimes$ denotes the Kronecker power. In other words, $G_n$ can be described recursively for $n\ge 4$ by the block matrix
\begin{equation*}
G_n=\left( \begin{array}{ccc}
G_{n/2} & 0\\
G_{n/2} & G_{n/2}\end{array}\right).
\end{equation*}
%

The matrix $G_n$ transforms $X_{[n]}$ into a random vector $U_{[n]}=X_{[n]}G_n$, such that the conditional entropy $H(U_i|U_{[i-1]},Y_{[n]})$ is \emph{polarized}. That means that for a fraction of close to $H(X|Y)$ of the indices $i\in[n]$, the conditional entropy $H(U_i|U_{[i-1]},Y_{[n]})$ is close to 1, and for almost all the rest of the indices, $H(U_i|U_{[i-1]},Y_{[n]})$ is close to 0. This result was shown by Arikan in~\cite{Ari09,Ari10}.
\begin{theorem}
\label{th:polarization}
\textbf{(Polarization Theorem)}
\cite[Theorem 1]{Ari10}
Let $n,U_{[n]},X_{[n]},Y_{[n]}$ be defined as above. For any $\delta\in(0,1)$, let 
$$H_{X|Y}\triangleq\mathset{i\in[n]:H(U_i|U_{[i-1]},Y_{[n]})\in(1-\delta,1)},$$
and 
$$L_{X|Y}\triangleq\mathset{i\in[n]:H(U_i|U_{[i-1]},Y_{[n]})\in(0,\delta)}.$$
Then 
$$\lim_{n\to\infty}|H_{X|Y}|/n=H(X|Y)\quad\text{and}\quad\lim_{n\to\infty}|L_{X|Y}|/n=1-H(X|Y).$$
\end{theorem}
Note that $H(X|Y)$ denotes a conditional entropy, while $H_{X|Y}$ denotes a subset of $[n]$. It is also shown in~\cite{Ari09} that the transformation $G_n$ is invertible with $G_n^{-1}=G_n$, implying $X_{[n]}=U_{[n]}G_n$. This polarization effect can be used quite simply for the design of a coding scheme that achieves the capacity of symmetric channels with a running time that is polynomial in the block length. The capacity of symmetric channels is achieved by a uniform distribution on the input alphabet, i.e. $p_X=1/2$~\cite[Theorem 7.2.1]{CovTho06}.
Since the input alphabet in this paper is binary, the capacity-achieving distribution gives $H(X)=1$, and therefore we have
\begin{equation}
\label{eq:sym_cap}
\lim_{n\to\infty}(1/n)|L_{X|Y}|=1-H(X|Y)=H(X)-H(X|Y)=I(X;Y)=C.
\end{equation}
Furthermore, for each index in $L_{X|Y}$, the conditional probability $p(u_i|u_{[i-1]},y_{[n]})$ must be close to either 0 or 1 (since the conditional entropy is small by the definition of the set $L_{X|Y}$). It follows that the RV $U_i$ can be estimated reliably given $u_{[i-1]}$ and $y_{[n]}$. This fact motivates the capacity-achieving coding scheme that follows. The encoder creates a vector $u_{[n]}$ by assigning the subvector $U_{L_{X|Y}}$ with the source message, and the subvector $U_{L_{X|Y}^c}$ with uniformly distributed random bits that are shared with the decoder. The randomness sharing is useful for the analysis, but is in fact unnecessary for using the scheme (the proof of this fact is described in~\cite[Section VI]{Ari09}). The set $U_{L_{X|Y}^c}$ is called the \emph{frozen set}. Equation~(\ref{eq:sym_cap}) implies that this coding rate approaches the channel capacity. The decoding is performed iteratively, from index 1 up to $n$. In each iteration, the decoder estimates the bit $u_i$ using the shared information or using a maximum likelihood estimation, according to the set membership of the iteration. The estimates of $u_i$ are denoted by $\hat{u}_i$. The estimates $\hat{u}_i$ for which $i$ is in $L_{X|Y}^c$ are always successful, since these bits were known to the decoder in advance. The rest of the bits (those in $L_{X|Y}$) are estimated correctly with high probability (as explained in the beginning of the paragraph), leading to a successful decoding of the entire message with high probability.

However, this reasoning does not translate directly to asymmetric channels. Remember that the capacity-achieving input distribution of asymmetric channels is in general not uniform (see, for example,~\cite{Gol80}), i.e. $p_X(1)\neq 1/2$. Since the Hadamard transform is bijective, it follows that the capacity-achieving distribution of the polarized vector $U_{[n]}$ is non uniform as well. The problem with this fact is that assigning uniform bits of message or shared randomness changes the distribution of $U_{[n]}$, and consequentially also changes the conditional entropies $H(U_i|U_{[i-1]},Y_{[n]})$.
To manage this situation, the approach proposed in~\cite{HonYam13}, which we adopt in this work, is to make sure that the change in the distribution of $U_{[n]}$ is kept to be minor, and thus its effect on the probability of decoding error is also minor.
To do this, consider the conditional entropies $H(U_i|U_{[i-1]})$, for $i\in[n]$. Since the polarization happens regardless of the channel model, we can consider a channel for which the output $Y$ is a deterministic variable, and conclude by Theorem~\ref{th:polarization} that the entropies $H(U_i|U_{[i-1]})$ also polarize. For this polarization, a fraction of $H(X)$ of the indices admit a high $H(U_i|U_{[i-1]})$. To ensure a minor change in the distribution of $U_{[n]}$, we restrict the assignments of uniform bits of message and shared randomness to the indices with high $H(U_i|U_{[i-1]})$. 

\begin{figure}[b]
\begin{center}
\includegraphics[width=0.3\linewidth]{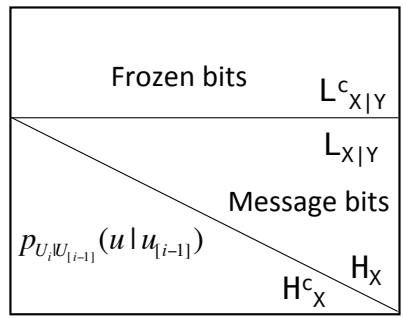}
\end{center}
\vspace{-10pt}
\caption{Encoding the vector $u_{[n]}$. The message bits are assigned in the set $H_X\cap L_{X|Y}$. The set $H_X^c\cap L_{X|Y}^c$ is vanishingly small, and shown as empty in the figure.}
\label{fig:asymmetric}
\vspace{-10pt}
\end{figure}

The insight of the last paragraph motivates a modified coding scheme. The locations with high entropy $H(U_i|U_{[i-1]})$ are assigned with uniformly distributed bits, while the rest of the locations are assigned with the pmf $p(u_i|u_{[i-1]})$. Note that $p(u_{[n]},x_{[n]})$ and $H(U_{[n]})$ refer to the capacity-achieving distribution of the channel, which does not equal to the distribution that the encoding process induces. Similar to the notation of Theorem~\ref{th:polarization}, we denote the set of indices with high entropy $H(U_i|U_{[i-1]})$ by $H_X$. To achieve a reliable decoding, we place the message bits in the indices of $H_X$ that can be decoded reliably, meaning that their entropies $H(U_i|U_{[i-1]},Y_{[n]})$ are low. So we say that we place the message bits in the intersection $H_X\cap L_{X|Y}$. 
The value of $u_{L_{X|Y}^c}$ must be known by the decoder in advance for a reliable decoding. Previous work suggested to share random Boolean functions between the encoder and the decoder, drawn according to the pmf $p(u_i|u_{[i-1]})$, and to assign the value of  $u_{(H_X\cap L_{X|Y})^c}=u_{H_X^c\cup L_{X|Y}^c}$ according to these functions \cite{GoeAbbGas13,HonYam13}. However, we note that the storage required for those Boolean functions is exponential in $n$, and therefore we propose an alternative method. 

To avoid the Boolean function, we divide the complement of $H_X\cap L_{X|Y}$ into three disjoint sets. First, the indices in the intersection $H_X\cap L_{X|Y}^c$ are assigned with uniformly distributed random bits that are shared between the encoder and the decoder. As in the symmetric case, this randomness sharing will in fact not be necessary, and a deterministic frozen vector could be shared instead. The rest of the bits of $U_{[n]}$ (those in the set $H_X^c$), are assigned \emph{randomly} at the encoder to a value $u$ with probability $p_{U_i|U_{[i-1]}}(u|u_{[i-1]})$ (where $p_{U_i|U_{[i-1]}}$ is calculated according to the pmf $p_{U_{[n]},X_{[n]},Y_{[n]}}$, the capacity-achieving distribution of the channel). The indices in $H_X^c\cap L_{X|Y}$ could be decoded reliably, but not those in $H_X^c\cap L_{X|Y}^c$. Fortunately, the set $H_X^c\cap L_{X|Y}^c$ can be shown to be small (as we will show later), and thus we could transmit those locations separately with a vanishing effect on the code rate. The encoding of the vector $u_{[n]}$ is illustrated in Figure~\ref{fig:asymmetric}.

We note that an alternative method to avoid the Boolean functions was in implied in~\cite{HonYam13}. According to this method, a seed of uniformly random bits is shared in advance between the encoder and the decoder. During encoding and decoding, the bits whose indices are in the set $(H_X\cap L_{X|Y})^c$ are generated as pseudorandom bits from the shared seed, such that each bit is distributed according to $p_{U_i|U_{[i-1]}}(u|u_{[i-1]})$. Such scheme could be used in many practical scenarios. However, the use of pseudorandomness might lead to error propagation in some applications. 
Therefore, we describe the constructions in the rest of the paper according to the approach of the previous paragraph.

To see the reason of why the code rate approaches the channel capacity, notice that the source message is placed in the indices in the intersection $H_X\cap L_{X|Y}$. The asymptotic fraction of this intersection can be derived as following.
\begin{equation}
\label{eq:rate}
|H_X\cap L_{X|Y}|/n=1-|H_X^c\cup L_{X|Y}^c|/n=1-|H_X^c|/n-|L_{X|Y}^c|/n+|H_X^c\cap L_{X|Y}^c|/n.
\end{equation}
The Polarization Theorem (Theorem~\ref{th:polarization}) implies that $|H_X^c|/n\to 1-H(X)$ and $|L_{X|Y}^c|/n\to H(X|Y)$. Since the fraction $|H_X^c\cap L_{X|Y}^c|$ vanishes for large $n$, we get that the asymptotic rate is $|H_X\cap L_{X|Y}|/n\to H(X)-H(X|Y)=I(X;Y)$, achieving the channel capacity.

For a more precise definition of the scheme, we use the so called Bhattacharyya parameter in the selection of subsets of $U_{[n]}$, instead of the conditional entropy. The Bhattacharyya parameters are polarized in a similar manner as the entropies, and are more useful for bounding the probability of decoding error. For a discrete RV $Y$ and a Bernoulli RV $X$, the Bhattacharyya parameter is defined by
\begin{equation}
\label{eq:bat_par}
Z(X|Y)\triangleq 2\sum_{y}\sqrt{p_{X,Y}(0,y)p_{X,Y}(1,y)}.
\end{equation}
Note that most of the polar coding literature is using a slightly different definition of the Bhattacharyya parameter, that coincides with Equation~(\ref{eq:bat_par}) when the RV $X$ is distributed uniformly. 
We use the following relations between the Bhattacharyya parameter and the conditional entropy.
\begin{proposition}
\label{prop:entropy}
(\cite[Proposition 2]{Ari10})
\begin{align}
(Z(X|Y))^2&\le H(X|Y),\\
H(X|Y)&\le\log_2(1+Z(X|Y))\le Z(X|Y).
\end{align}
\end{proposition}
We now define the set of high and low Bhattacharyya parameters, and work with them instead of the sets $H_{X|Y}$ and $L_{X|Y}$. For $\delta\in (0,1)$, define
\begin{align*}
\mathcal{H}_{X|Y}&\triangleq\mathset{i\in[n]:Z(U_i|U_{[i-1]},Y_{[n]})\ge 1-2^{-n^{1/2-\delta}}},\\
\mathcal{L}_{X|Y}&\triangleq\mathset{i\in[n]:Z(U_i|U_{[i-1]},Y_{[n]})\le 2^{-n^{1/2-\delta}}}.
\end{align*}
As before, we define the sets $\mathcal{H}_{X}$ and $\mathcal{L}_{X}$ for the parameter $Z(U_i|U_{[i-1]})$ by letting $Y_{[n]}$ be a deterministic vector. Using Proposition~\ref{prop:entropy}, it is shown in~\cite[combining Proposition 2 with Theorem 2]{HonYam13} that Theorem~\ref{th:polarization} holds also if we replace the sets  ${H}_{X|Y}$ and ${L}_{X|Y}$ with the sets $\mathcal{H}_{X|Y}$ and $\mathcal{L}_{X|Y}$. That is, we have
\begin{equation}
\label{eq:sizes}
\lim_{n\to\infty}|\mathcal{H}_{X|Y}|/n=H(X|Y)\quad\text{and}\quad\lim_{n\to\infty}|\mathcal{L}_{X|Y}|/n=1-H(X|Y).
\end{equation}

We now define our coding scheme formally. Let $m_{[|\mathcal{H}_X\cap\mathcal{L}_{X|Y}|]}\in\mathset{0,1}^{|\mathcal{H}_X\cap\mathcal{L}_{X|Y}|}$ be the realization of a uniformly distributed source message, and $f_{[|\mathcal{H}_X\cap\mathcal{L}_{X|Y}^c|]}\in\mathset{0,1}^{|\mathcal{H}_X\cap\mathcal{L}_{X|Y}^c|}$ be a deterministic frozen vector known to both the encoder and the decoder. We discuss how to find a good frozen vector in Appendix~\ref{app:existance}. For a subset $\mathcal{A}\subseteq [n]$ and an index $i\in\mathcal{A}$, we use a function $r(i,\mathcal{A})$ to denote the \emph{rank} of $i$ in an ordered list of the elements of $\mathcal{A}$. 
The probabilities $p_{U_i|U_{[i-1]}}(u|u_{[i-1]})$ and $p_{U_i|U_{[i-1]},Y^n}(u|u_{[i-1]},y_{[n]})$ can be calculated efficiently by a recursive method described in~\cite[Section III.B]{HonYam13}.

\begin{construction}
\label{con:asymmetric}
\\
\noindent \textbf{Encoding}

\textbf{Input}: a message $m_{[|\mathcal{H}_X\cap\mathcal{L}_{X|Y}|]}\in\mathset{0,1}^{|\mathcal{H}_X\cap\mathcal{L}_{X|Y}|}$.

\textbf{Output:} a codeword $x_{[n]}\in\mathset{0,1}^n$.

\begin{enumerate}
\item For $i$ from $1$ to $n$, successively, set
\begin{equation*}
\label{eq:as_encoding}
u_i= \left\{ 
  \begin{array}{l l}
		u\in\mathset{0,1} \quad \text{with probability $p_{U_i|U_{[i-1]}}(u|u_{[i-1]})$}& \quad \text{} \text{if $i\in\mathcal{H}_{X}^c$}\\
	  m_{r(i,\mathcal{H}_{X}\cap\mathcal{L}_{X|Y})}& \quad \text{if $i\in\mathcal{H}_{X}\cap\mathcal{L}_{X|Y}$}\\
	  f_{r(i,\mathcal{H}_{X}\cap\mathcal{L}_{X|Y}^c)} & \quad \text{if $i\in\mathcal{H}_{X}\cap\mathcal{L}_{X|Y}^c$.}
  \end{array} \right.
\end{equation*}
\item Transmit the codeword $x_{[n]}=u_{[n]}G_{n}$.
\item Transmit the vector $u_{\mathcal{H}_{X}^c\cap\mathcal{L}_{X|Y}^c}$ separately using a linear, non-capacity-achieving polar code with a uniform input distribution (as in~\cite{Ari09}). In practice, other error-correcting codes could be used for this vector as well.

\end{enumerate}

\noindent \textbf{Decoding}

\textbf{Input}: a noisy vector $y_{[n]}\in\mathset{0,1}^n$.

\textbf{Output:} a message estimation $\hat{m}_{[|\mathcal{H}_X\cap\mathcal{L}_{X|Y}|]}\in\mathset{0,1}^{|\mathcal{H}_X\cap\mathcal{L}_{X|Y}|}$.

\begin{enumerate}

\item Estimate the vector $u_{\mathcal{H}_{X}^c\cap\mathcal{L}_{X|Y}^c}$ by $\hat{u}_{\mathcal{H}_{X}^c\cap\mathcal{L}_{X|Y}^c}$.

\item For $i$ from 1 to $n$, set
\begin{equation*}
\label{eq:as_decoding}
\hat{u}_i= \left\{ 
  \begin{array}{l l}
    \displaystyle\argmax_{u\in\mathset{0,1}}p_{U_i|U_{[i-1]},Y_{[n]}}(u|u_{[i-1]},y_{[n]})& \quad \text{if $i\in \mathcal{L}_{X|Y}$}\\
		\hat{u}_{r(i,\mathcal{H}_{X}^c\cap\mathcal{L}_{X|Y}^c)} &\quad \text{if $i\in \mathcal{H}_{X}^c\cap\mathcal{L}_{X|Y}^c$}\\
		f_{r(i,\mathcal{H}_{X}\cap\mathcal{L}_{X|Y}^c)} & \quad \text{if $i\in \mathcal{H}_{X}\cap\mathcal{L}_{X|Y}^c$.}
  \end{array} \right.
\end{equation*}

\item Return the estimated message $\hat{m}_{[|\mathcal{H}_{X}\cap\mathcal{L}_{X|Y}|]}=\hat{u}_{\mathcal{H}_{X}\cap\mathcal{L}_{X|Y}}$.
\end{enumerate}
\end{construction}

We say that a sequence of coding schemes achieves the channel capacity if the probability of decoding error vanishes with the block length for any rate below the capacity. 
\begin{theorem}
\label{th:asymmetric}
Construction~\ref{con:asymmetric} achieves the channel capacity (Theorem~\ref{th:capacity}) with a  encoding and decoding complexity of $O(n\log n)$  and a probability of decoding error of at most $2^{-n^{1/2-\delta}}$ for any $\delta>0$ and large enough $n$.
\end{theorem}
In the next section we show a generalized construction and prove its capacity-achieving property. Theorem~\ref{th:asymmetric} thus will follow as a corollary of the more general Theorem~\ref{th:multicoding}.
We note here two differences between Construction~\ref{con:asymmetric} and the construction in~\cite[Section III.B]{MonHasSasUrb14}. First, in the encoding of Construction~\ref{con:asymmetric}, the bits in the set $\mathcal{H}_X^c$ are set randomly, while in~\cite[Section III.B]{MonHasSasUrb14}, those bits are set according to a maximum likelihood rule. And second, the vector $u_{\mathcal{H}_{X}^c\cap\mathcal{L}_{X|Y}^c}$ is being sent through a side channel in Construction~\ref{con:asymmetric}, but not in~\cite[Section III.B]{MonHasSasUrb14}. 
These two features of Construction~\ref{con:asymmetric} allow an alternative analysis and proof that the scheme achieves the channel capacity. 


\section{Channels with Non-Causal Encoder State Information}
\label{sec:multicoding}

In this section we generalize Construction~\ref{con:asymmetric} to the availability of channel state information at the encoder. We consider mainly the application of rewriting in flash memories, and present two special cases of the channel model for this application. In flash memory, information is stored in a set of $n$ memory cells. We mainly focus on a flash memory type that is called Single-Level Cell (SLC), in which each cell stores a single information bit, and its value is denoted by either 0 or 1. We first note that the assumption of a memoryless channel is not exactly accurate in flash memories, due to a mechanism of cell-to-cell interference. However, we keep using this assumption, as it is nonetheless useful for the design of coding schemes with valuable practical performance. The main limitation of flash memories that we consider in this work is the high cost of changing a cell level from 1 to 0 (in SLC memories). To perform such a change, an expensive operation, called ``block erasure'', is required. To avoid this block erasure operation, information is rewritten over existing memory in the sense that no cell is changed from value 1 to 0. We thus consider the use of the information about the previous state of the cells in the encoding process. 
We model the memory cells as a channel with a discrete \emph{state}, and we also assume that the state is memoryless, meaning that the states of different cells are distributed independently. 
 
We assume that the state of the entire $n$ cells is available to the writer prior to the beginning of the writing process. In communication terminology this kind of state availability is refereed to as ``non causal''. We note that this setting is also useful in the so called \emph{Marton-coding} method for communication over broadcast channels. Therefore, the multicoding schemes that will follow serve as a contribution also in this important setting. One special case of the model which we consider is the noiseless write-once memory model. This model also serves as an ingredient for a type of codes called ``rank-modulation rewriting codes''~\cite{EngYaaJiaBru13}. Therefore, the schemes proposed in this section can also be useful for the design of rank-modulation rewriting codes.

We represent the channel state as a Bernoulli random variable $S$ with parameter $\beta$, which equals the probability $p_S(S=1)$. A cell of state 1 can only be written with the value 1. Note that, intuitively, when $\beta$ is high, the capacity of the memory is small, since only a few cells are available for modification in the writing process, and thus only a small amount of information could be stored. This also means that the choice of codebook has a crucial effect on the capacity of the memory in \emph{future writes}. A codebook that contains many codewords of high Hamming weight (number of 1's in the codeword) would make the parameter $\beta$ of future writes high, and thus the capacity of the future writes would be low. However, forcing the expected Hamming weight of the codebook to be low would reduce the capacity of the current write. To settle this trade-off, previous work suggested to optimize the sum of the code rates over multiple writes. It was shown that in many cases, constraints on the codebook Hamming weight (henceforth just weight) strictly increase the sum rate (see, for example,~\cite{Hee85}). Therefore, we consider an input cost constraint in the model. 

The most general model that we consider is a discrete memoryless channel (DMC) with a discrete memoryless (DM) state and an input cost constraint, where the state information is available non causally at the encoder. The channel input, state and output are denoted by $x,s$ and $y$, respectively, and their respective finite alphabets are denoted by $\mathcal{X},\mathcal{S}$ and $\mathcal{Y}$. The random variables are denoted by $X,S$ and $Y$, and the random vectors by $X_{[n]},S_{[n]}$ and $Y_{[n]}$, where $n$ is the block length. The state is distributed according to the pmf $p(s)$, and the conditional pmfs of the channel are denoted by $p(y|x,s)$. The input cost function is denoted by $b(x)$, and the input cost constraint is 
$$\sum_{i=1}^{n}\mathsf{E}[b(X_i)]\le nB,$$
where $B$ is a real number representing the normalized constraint.
The channel capacity with an informed encoder and an input cost constraint is given by an extension of the Gelfand-Pinsker Theorem\footnote{The cost constraint is defined slightly differently in this reference, but the capacity is not affected by this change.}.
\begin{theorem}
\label{th:gp_capacity}
\textbf{(Gelfand-Pinsker Theorem with Cost Constraint)}
\cite[Equation (7.7) on p. 186]{ElgKim12}
Consider a DMC with a DM state where $p(y|x,s)$ denotes the channel transition probability and $p(s)$ denotes the state probability. Under an input cost constraint $B$, where the state information is available non causally only at the encoder, the capacity of the channel is
\begin{equation}
\label{eq:capacity}
C=\max_{p(v|s),x(v,s):\mathsf{E}(b(X))\le B}(I(V;Y)-I(V;S)),
\end{equation}
where $V$ is an auxiliary random variable with a finite alphabet $\mathcal{V}$, $|\mathcal{V}|\le\min\mathset{|\mathcal{X}\cdot\mathcal{S},\mathcal{Y}+\mathcal{S}-1}$.
\end{theorem}

The main coding scheme that we present in this section achieves the capacity in Theorem~\ref{th:gp_capacity}. The proof of Theorem~\ref{th:gp_capacity} considers a virtual channel model, in which the RV $V$ is the channel input and $Y$ is the channel output. Similar to the previous section, we limit the treatment to the case in which the RV $V$ is binary. In flash memory, this case would correspond to a single-level cell (SLC) type of memory. As mentioned in Section~\ref{sec:asymmetric}, an extension of the scheme to a non-binary case is not difficult. The non-binary case is useful for flash memories in which each cell stores 2 or more bits of information. Such memories are called Multi-Level Cell (MLC). We also mention that the limitation to binary random variables does not apply on the channel output $Y$. Therefore, the cell voltage in flash memory could be read more accurately at the decoder to increase the coding performance, similarly to the soft decoding method that is used in flash memories with LDPC codes. Another practical remark is that the binary-input model can be used in MLC memories by coding separately on the MSB and the LSB of the cells, as in fact is the coding method in current MLC flash systems.

The scheme that achieves the capacity of Theorem~\ref{th:gp_capacity} is called  Construction~\ref{con:chaining}, and it will be described in Subsection~\ref{sub:chaining}. The capacity achieving result is summarized in the following theorem, which will be proven in Subsection~\ref{sub:chaining}.
\begin{theorem}
\label{th:chaining}
Construction~\ref{con:chaining} achieves the capacity of the Gelfand-Pinsker Theorem with Cost Constraint (Theorem~\ref{th:gp_capacity}) with a encoding and decoding complexity of $O(n\log n)$  and a probability of decoding error of at most $2^{-n^{1/2-\delta}}$ for any $\delta>0$ and large enough $n$.
\end{theorem}

Note that the setting of Theorem~\ref{th:chaining} is a generalization of the asymmetric channel-coding setting of Theorem~\ref{th:asymmetric}, and therefore Construction~\ref{con:chaining} and Theorem~\ref{th:chaining} are in fact a generalization of Construction~\ref{con:asymmetric} and Theorem~\ref{th:asymmetric}.
We note also that polar codes were constructed for a symmetric case of the Gelfand-Pinsker channel by Korada and Urbanke in~\cite{KorUrb10}. As the key constraint of flash memories is notably asymmetric, the important novelty of this work is in providing the non-trivial generalization that cover the asymmetric case.

Before we describe the code construction, we first show in Subsection~\ref{sub:special} two special cases of the Gelfand-Pinsker model that are useful for the rewriting of flash memories. Afterwards, in subsections~\ref{sub:degraded} and~\ref{sub:chaining}, we will show two versions of the construction that correspond to generalizations of the two special cases.

\subsection{Special Cases}
\label{sub:special}

We start with a special case that is quite a natural model for flash memory rewriting.

\begin{figure}[t]
\begin{center}
\includegraphics[width=0.51\linewidth]{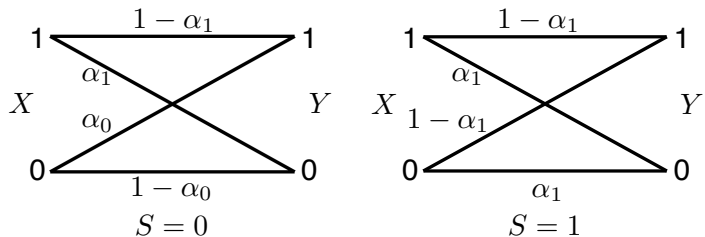}
\end{center}
\vspace{-10pt}
\caption{Example~\ref{ex:realistic}: A binary noisy WOM model.}
\label{fig:realistic}
\vspace{-5pt}
\end{figure}

\begin{example}
\label{ex:realistic}
Let the sets $\mathcal{X},\mathcal{S}$ and $\mathcal{Y}$ be all equal to $\mathset{0,1}$, and let the state pmf be  $p_S(1)=\beta$. This model corresponds to a single level cell flash memory. We describe the cell behaviour after a bit $x$ is attempted to be written. When $s=0$, the cell behaves as a binary asymmetric channel with input $x$, since the call state does not interfere with the writing attempt. When $s=1$, the cell behaves as if a value of 1 was attempted to be written, regardless of the actual value $x$ attempted. However, an error might still occur, during the writing process or anytime afterwards (for example, due to charge leakage). Thus, we can say that when $s=1$, the cell behaves as a binary asymmetric channel with input 1. Formally, the channel pmfs are given by 
\begin{equation}
\label{eq:transition_accurate}
p_{Y|XS}(1|x,s)= \left\{ 
  \begin{array}{l l}
    \alpha_0 & \quad \text{if $(x,s)=(0,0)$}\\
    1-\alpha_1 & \quad \text{if $(x,s)=(0,1)$}\\
		1-\alpha_1 & \quad \text{if $(x,s)=(1,0)$}\\
    1-\alpha_1 & \quad \text{if $(x,s)=(1,1)$}\\
  \end{array} \right.
\end{equation}

The error model is also presented in Figure~\ref{fig:realistic}. The cost constraint is given by $b(x_i)=x_i$, since it is desirable to limit the amount of cells written to a value of 1.
\end{example}

Our coding-scheme construction for the setting of Theorem~\ref{th:gp_capacity} is based on a more limited construction, which serves as a building block. We will start by describing the limited construction, and then show how to extend it for the model of Theorem~\ref{th:gp_capacity}. We will prove that the limited construction achieves the capacity of channels whose capacity-achieving distribution forms a certain stochastically degraded structure. We first recall the definition of stochastically degraded channels.

\begin{definition}
\label{def:degradation}
\cite[p. 112]{ElgKim12} A discrete memoryless channel (DMC) $W_1:\mathset{0,1}\to\mathcal{Y}_1$  is  \emph{stochastically degraded} (or simply degraded) with respect to a DMC $W_2:\mathset{0,1}\to\mathcal{Y}_2$, denoted as $W_1\preceq W_2$, if there exists a DMC $W:\mathcal{Y}_2\to\mathcal{Y}_1$ such that $W$ satisfies the equation $W_1(y_1|x)=\sum_{y_2\in\mathcal{Y}_2}W_2(y_2|x)W(y_1|y_2)$.
\end{definition}
Next, we bring the required property of channels whose capacity is achieved by the limited construction to be proposed.

\begin{property}
\label{pro:degradation}
There exist functions $p(v|s)$ and $x(v,s)$ that maximize the Gelfand-Pinsker capacity in Theorem~\ref{th:gp_capacity} which satisfy the condition $p(y|v)\succeq p(s|v)$.
\end{property}

\begin{figure}[t]
\begin{center}
\includegraphics[width=0.51\linewidth]{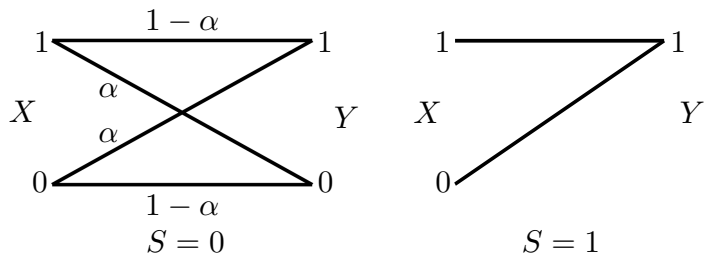}
\end{center}
\vspace{-10pt}
\caption{Example~\ref{ex:degraded}: A binary WOM with writing noise.}
\label{fig:degraded}
\vspace{-5pt}
\end{figure}

It is an open problem whether the model of Example~\ref{ex:realistic} satisfies the degradation condition of Property~\ref{pro:degradation}. However, we can modify the model such that it will satisfy Property~\ref{pro:degradation}. Specifically, we study the following model:
\begin{example}
\label{ex:degraded}
Let the sets $\mathcal{X},\mathcal{S}$ and $\mathcal{Y}$ be all equal to $\mathset{0,1}$. The channel and state pmfs are given by $p_S(1)=\beta$ and 
\begin{equation}
\label{eq:channel_given_state}
p_{Y|XS}(1|x,s)= \left\{ 
  \begin{array}{l l}
    \alpha & \quad \text{if $(x,s)=(0,0)$}\\
    1-\alpha & \quad \text{if $(x,s)=(1,0)$}\\
		1 & \quad \text{if $s=1$}.
  \end{array} \right.
\end{equation}
In words, if $s=1$ the channel output is always 1, and if $s=0$, the channel behave as a binary symmetric channel. The cost function is given by $b(x_i)=x_i$. The error model is also presented in Figure~\ref{fig:degraded}. This model can represent a writing noise, as a cell of state $s=1$ is not written on and it never suffers errors.
\end{example}

We claim that the model of Example~\ref{ex:degraded} satisfies the degradation condition of Property~\ref{pro:degradation}. To show this, we need first to find the functions $p(v|s)$ and $x(v,s)$ that maximize the Gelfand-Pinsker capacity in Theorem~\ref{th:gp_capacity}. Those functions are established in the following theorem of Heegard.

\begin{theorem}
\label{th:degraded_capacity}\cite[Theorem 4]{Hee85}
The capacity of the channel in Example~\ref{ex:degraded} is 
$$C=(1-\beta)[h(\epsilon*\alpha)-h(\alpha)],$$
where $\epsilon=B/(1-\beta)$ and $\epsilon*\alpha\equiv \epsilon(1-\alpha)+(1-\epsilon)\alpha$.
The selections $\mathcal{V}=\mathset{0,1}$, $x(v,s)=v\wedge \neg s$ (where $\wedge$ is the logical AND operation, and $\neg$ is the logical negation),  and
\begin{equation}
\label{eq:cond_dist}
p_{V|S}(1|0)=\epsilon, \qquad p_{V|S}(1|1)=\frac{\epsilon(1-\alpha)}{\epsilon*\alpha}
\end{equation}
achieve this capacity.
\end{theorem}

We provide an alternative proof for Theorem~\ref{th:degraded_capacity} in Appendix~\ref{app:capacity}. Intuitively, the upper bound is obtained by assuming that the state information is available also at the decoder, and the lower bound is obtained by setting the functions $x(v,s)$ and $p(v|s)$ according to the statement of the theorem. The proof that the model in Example~\ref{ex:degraded} satisfies the degradation condition of Property~\ref{pro:degradation} is completed by the following lemma.
\begin{lemma}
\label{lem:degradation}
The capacity achieving functions of Theorem~\ref{th:degraded_capacity} for the model of Example~\ref{ex:degraded} satisfy the degradation condition of Property~\ref{pro:degradation}. That is, the channel $p(s|v)$ is degraded with respect to the channel $p(y|v)$.
\end{lemma}
Lemma~\ref{lem:degradation} is proven in Appendix \ref{app:degradation}, and consequently, the capacity of the model in Example~\ref{ex:degraded} can be achieved by our limited construction. In the next subsection we describe the construction for channel models which satisfy Property~\ref{pro:degradation}, including the model in Example~\ref{ex:degraded}.

\subsection{Multicoding Construction for Degraded Channels}
\label{sub:degraded}

Notice first that the capacity-achieving distribution of the asymmetric channel in Section~\ref{sec:asymmetric} actually satisfies Property~\ref{pro:degradation}. In the asymmetric channel-coding case, the state can be thought of as a degenerate random variable (a RV which only takes a single value), and therefore we can choose $W$ in Definition~\ref{def:degradation} to be degenerate as well, and by that satisfy Property~\ref{pro:degradation}. We will see that the construction that we present in this subsection is a generalization of Construction~\ref{con:asymmetric}. 

\begin{figure}[b]
\begin{center}
\includegraphics[width=0.3\linewidth]{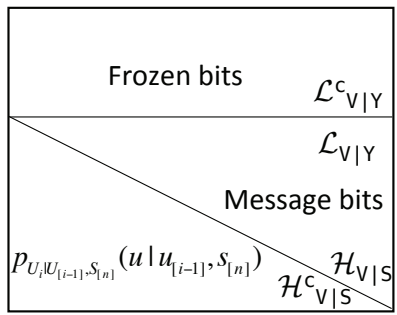}
\end{center}
\vspace{-10pt}
\caption{Encoding the vector $u_{[n]}$ in Construction~\ref{con:multicoding}. The message bits are assigned in the set $\mathcal{H}_{V|S}\cap \mathcal{L}_{V|Y}$. The set $\mathcal{H}_{V|S}^c\cap \mathcal{L}_{V|Y}^c$ is vanishingly small, and shown as empty in the figure.}
\label{fig:degraded_sets}
\vspace{-10pt}
\end{figure}

The construction has a similar structure as the achievability proof of the Gelfand-Pinsker Theorem (Theorem~\ref{th:gp_capacity}). The encoder first finds a vector $v_{[n]}$ in a similar manner to Construction~\ref{con:asymmetric}, where the RV $X|Y$ in Construction~\ref{con:asymmetric} is replaced with $V|Y$, and the RV $X$ is replaced with $V|S$. The vector $U_{[n]}$ is now the polarization of the vector $V_{[n]}$, meaning that $U_{[n]}=V_{[n]}G_n$. The RV $V$ is taken according to the pmfs $p(v|s)$ that maximize the rate expression in Equation~(\ref{eq:capacity}). The selection of the vector $u_{[n]}$ is illustrated in Figure~\ref{fig:degraded_sets}. After the vector $u_{[n]}$ is chosen, each bit $i\in[n]$ in the codeword $x_{[n]}$ is calculated by the function $x_i(v_i,s_i)$ that maximizes Equation~(\ref{eq:capacity}). To use the model of Example~\ref{ex:degraded}, one should use the functions $p(v|s)$ and $x(v,s)$ according to Theorem~\ref{th:degraded_capacity}. The key to showing that the scheme achieves the channel capacity is that the fraction $|\mathcal{H}_{V|S}^c\cap \mathcal{L}_{V|Y}^c|/n$ can be shown to vanish for large $n$ if the channel satisfies Property~\ref{pro:degradation}. Then, by the same intuition as in Equation~(\ref{eq:rate}) and using Equation~(\ref{eq:sizes}), the replacements imply that the asymptotic rate of the codes is %

\begin{align*}
\label{eq:gp_rate}
|\mathcal{H}_{V|S}\cap\mathcal{L}_{V|Y}|/n&=1-|\mathcal{H}_{V|S}^c|/n-|\mathcal{L}_{V|Y}^c|/n+|\mathcal{H}_{V|S}^c\cap \mathcal{L}_{V|Y}^c|/n\\
&\to 1-(1-H(V|S))-H(V|Y)+0\\
&=I(V;Y)-I(V;S),
\end{align*}
achieving the Gelfand-Pinsker capacity of Theorem~\ref{th:gp_capacity}. We now describe the coding scheme formally. 
\begin{construction}
\label{con:multicoding}
\\
\noindent \textbf{Encoding}

\textbf{Input}: a message $m_{[|\mathcal{H}_{V|S}\cap\mathcal{L}_{V|Y}|]}\in\mathset{0,1}^{|\mathcal{H}_{V|S}\cap\mathcal{L}_{V|Y}|}$ and a state $s_{[n]}\in\mathset{0,1}^n$.

\textbf{Output:} a codeword $x_{[n]}\in\mathset{0,1}^n$.

\begin{enumerate}
\item For each $i$ from $1$ to $n$, assign
\begin{equation}
\label{eq:encoding}
u_i= \left\{ 
  \begin{array}{l l}
		u\in\mathset{0,1} \quad \text{with probability $p_{U_i|U_{[i-1]},S_{[n]}}(u|u_{[i-1]},s_{[n]})$}& \quad \text{} \text{if $i\in\mathcal{H}_{V|S}^c$}\\
	  m_{r(i,\mathcal{H}_{V|S}\cap\mathcal{L}_{V|Y})}& \quad \text{if $i\in\mathcal{H}_{V|S}\cap\mathcal{L}_{V|Y}$}\\
		f_{r(i,\mathcal{H}_{V|S}\cap\mathcal{L}_{V|Y}^c)} & \quad \text{if $i\in\mathcal{H}_{V|S}\cap\mathcal{L}_{V|Y}^c$.}
  \end{array} \right.
\end{equation}
\item Calculate $v_{[n]}=u_{[n]}G_{n}$ and for each $i\in [n]$, store the value $x_i(v_i,s_i)$.
\item Store the vector $u_{\mathcal{H}_{V|S}^c\cap\mathcal{L}_{V|Y}^c}$ separately using a point-to-point linear non-capacity-achieving polar code with a uniform input distribution. The encoder here does not use the state information in the encoding process, but rather treat it as an unknown part of the channel noise.
\end{enumerate}

\noindent \textbf{Decoding}

\textbf{Input}: a noisy vector $y_{[n]}\in\mathset{0,1}^n$.

\textbf{Output:} a message estimation $\hat{m}_{[|\mathcal{H}_{V|S}\cap\mathcal{L}_{V|Y}|]}\in\mathset{0,1}^{|\mathcal{H}_{V|S}\cap\mathcal{L}_{V|Y}|}$.

\begin{enumerate}

\item Estimate the vector $u_{\mathcal{H}_{V|S}^c\cap\mathcal{L}_{V|Y}^c}$ by $\hat{u}_{\mathcal{H}_{V|S}^c\cap\mathcal{L}_{V|Y}^c}$.

\item Estimate $u_{[n]}$ by $\hat{u}_{[n]}(y_{[n]},f_{[|\mathcal{H}_{V|S}\cap\mathcal{L}_{V|Y}^c|]})$ as follows: For each $i$ from 1 to $n$, assign
\begin{equation}
\label{eq:multidecoding}
\hat{u}_i= \left\{ 
  \begin{array}{l l}
    \displaystyle\argmax_{u\in\mathset{0,1}}p_{U_i|U_{[i-1]},Y_{[n]}}(u|u_{[i-1]},y_{[n]})& \quad \text{if $i\in \mathcal{L}_{V|Y}$}\\
		\hat{u}_{r(i,\mathcal{H}_{V|S}^c\cap\mathcal{L}_{V|Y}^c)} &\quad \text{if $i\in \mathcal{H}_{V|S}^c\cap\mathcal{L}_{V|Y}^c$}\\
		f_{r(i,\mathcal{H}_{V|S}\cap\mathcal{L}_{V|Y}^c)} & \quad \text{if $i\in \mathcal{H}_{V|S}\cap\mathcal{L}_{V|Y}^c$.}
  \end{array} \right.
\end{equation}
\item Return the estimated message $\hat{m}_{[|\mathcal{H}_{V|S}\cap\mathcal{L}_{V|Y}|]}=\hat{u}_{\mathcal{H}_{V|S}\cap\mathcal{L}_{V|Y}}$.
\end{enumerate}
\end{construction}

The asymptotic performance of Construction \ref{con:multicoding} is stated in the following theorem.
\begin{theorem}
\label{th:multicoding}
If Property~\ref{pro:degradation} holds, then Construction~\ref{con:multicoding} achieves the capacity of Theorem~\ref{th:gp_capacity} with a encoding and decoding complexity of $O(n\log n)$  and a probability of decoding error of at most $2^{-n^{1/2-\delta}}$ for any $\delta>0$ and large enough $n$.
\end{theorem}

The proof of Theorem~\ref{th:multicoding} is shown in Appendix~\ref{app:multicoding}. The next subsection describes a method to remove the degradation requirement of Property~\ref{pro:degradation}. This allows to achieve also the capacity of the more realistic model of Example~\ref{ex:realistic}.

\subsection{Multicoding Construction without Degradation}
\label{sub:chaining}

A technique called ``chaining'' was proposed in~\cite{MonHasSasUrb14} that allows to achieve the capacity of models that do not exhibit the degradation condition of Property~\ref{pro:degradation}. The chaining idea was presented in the context of broadcast communication and point-to-point universal coding. We connect it here to the application of flash memory rewriting through Example~\ref{ex:realistic}. We note also that the chaining technique that follows comes with a price of a slower convergence to the channel capacity, and thus a lower non-asymptotic code rate. 

The requirement of Construction~\ref{con:multicoding} for degraded channels comes from the fact that the set $\mathcal{H}_{V|S}^c\cap\mathcal{L}_{V|Y}^c$ needs to be communicated to the decoder in a side channel.
If the fraction $(1/n)|\mathcal{H}_{V|S}^c\cap\mathcal{L}_{V|Y}^c|$ vanishes with $n$, Construction~\ref{con:multicoding} achieves the channel capacity. In this subsection we deal with the case that the fraction  $(1/n)|\mathcal{H}_{V|S}^c\cap\mathcal{L}_{V|Y}^c|$ does not vanish. In this case we have
\begin{align*}
\label{eq:set_size}
|H_{V|S}\cap L_{V|Y}|/n=&1-|H_{V|S}^c\cup L_{V|Y}^c|/n\\
=&1-|H_{V|S}^c|/n-|L_{V|Y}^c|/n+|H_{V|S}^c\cap L_{V|Y}^c|/n\\
\to&I(V;Y)-I(V;S)+|H_{V|S}^c\cap L_{V|Y}^c|/n.
\end{align*}

\begin{figure}[b]
\begin{center}
\includegraphics[width=1\linewidth]{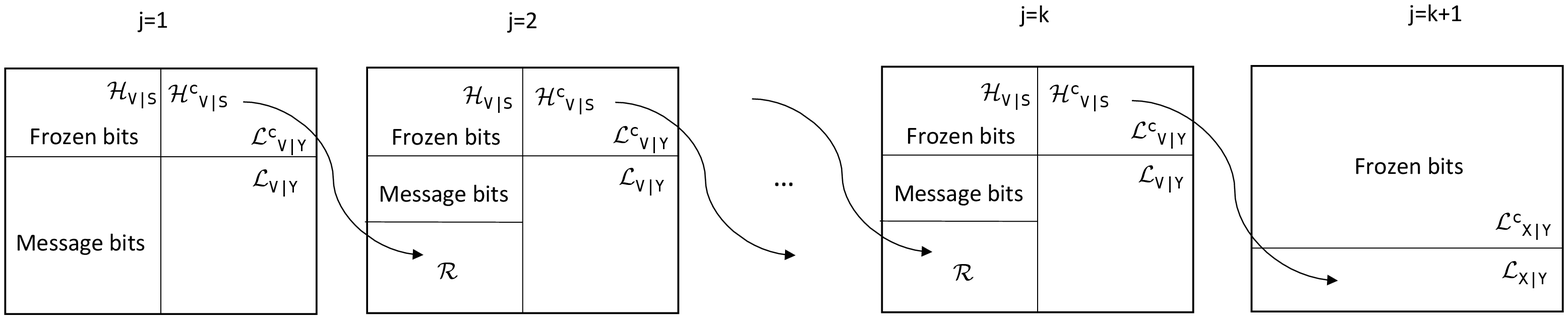}
\end{center}
\vspace{-10pt}
\caption{The chaining construction}
\label{fig:chaining}
\vspace{-5pt}
\end{figure}

The idea is then to store the subvector $u_{\mathcal{H}_{V|S}^c\cap\mathcal{L}_{V|Y}^c}$ in a subset of the indices $\mathcal{H}_{V|S}\cap\mathcal{L}_{V|Y}$ of an \emph{additional code block} of $n$ cells. The additional block is using the same coding technique as the original block. Therefore, it can use about $I(V;Y)-I(V;S)$ of the cells to store additional message bits, and by that to approach the channel capacity. We denote the by $\mathcal{R}$ the subset of $\mathcal{H}_{V|S}\cap\mathcal{L}_{V|Y}$ in which we store the the subvector $u_{\mathcal{H}_{V|S}^c\cap\mathcal{L}_{V|Y}^c}$ of the previous block. Note that the additional block also faces the same difficulty as the original block with the set $H_{V|S}^c\cap L_{V|Y}^c$. To solve this, we use the same solution, recursively, sending a total of $k$ blocks, each of length $n$. Each block can store a source message of fraction that approaches the channel capacity. The ``problematic'' bits of block $k$ (the last block) will then be stored using yet another block, but this block will be coded without taking the state information into account, and thus will not face the same difficulty. The last block is thus causing a rate loss, but this loss is of a fraction $1/k$, which vanishes for large $k$. The decoding is performed ``backwards'', starting from the last block and ending with the first block. The chaining construction is illustrated in Figure~\ref{fig:chaining}.
In the following formal description of the construction we denote the index $i$ of the $j$-th block of the message by $m_{i,j}$, and similarly for other vectors. The vectors themselves are are also denoted in two dimensions, for example $x_{[n],[k]}$.

\begin{construction}
\label{con:chaining}
\\
Let $\mathcal{R}$ be an arbitrary subset of $\mathcal{H}_{V|S}\cap\mathcal{L}_{V|Y}$ of size $|\mathcal{H}_{V|S}^c\cap\mathcal{L}_{V|Y}^c|$.

\noindent \textbf{Encoding}

\textbf{Input}: a message $m_{([|\mathcal{H}_{V|S}\cap\mathcal{L}_{V|Y}|-|\mathcal{H}_{V|S}^c\cap\mathcal{L}_{V|Y}^c|],[k])}\in\mathset{0,1}^{k(|\mathcal{H}_{V|S}\cap\mathcal{L}_{V|Y}|-|\mathcal{H}_{V|S}^c\cap\mathcal{L}_{V|Y}^c|)}$ and a state $s_{[n],[k]}\in\mathset{0,1}^{kn}$.

\textbf{Output:} a codeword $x_{[n],[k]}\in\mathset{0,1}^{kn}$.

\begin{enumerate}
\item Let $u_{[n],0}\in\mathset{0,1}^n$ be an arbitrary vector. For each $j$ from $1$ to $k$, and for each $i$ from $1$ to $n$, assign
\begin{equation}
\label{eq:chaining_encoding}
u_{i,j}= \left\{ 
  \begin{array}{l l}
		u\in\mathset{0,1} \quad \text{with probability $p_{U_i|U_{[i-1]},S_{[n]}}(u|u_{[i-1]},s_{[n]})$}& \quad \text{} \text{if $i\in\mathcal{H}_{V|S}^c$}\\
	  m_{r(i,\mathcal{H}_{V|S}\cap\mathcal{L}_{V|Y}),j}& \quad \text{if $i\in(\mathcal{H}_{V|S}\cap\mathcal{L}_{V|Y})\setminus\mathcal{R}$}\\
		u_{r(i,\mathcal{H}_{V|S}^c\cap\mathcal{L}_{V|Y}^c),j-1}& \quad \text{if $i\in\mathcal{R}$}\\
		f_{r(i,\mathcal{H}_{V|S}\cap\mathcal{L}_{V|Y}^c),j} & \quad \text{if $i\in\mathcal{H}_{V|S}\cap\mathcal{L}_{V|Y}^c$.}
  \end{array} \right.
\end{equation}
\item For each $j$ from $1$ to $k$ calculate $v_{[n],j}=u_{[n],j}G_{n}$, and for each $i\in [n]$, store the value $x_{i,j}(v_{i,j},s_{i,j})$.
\item Store the vector $u_{\mathcal{H}_{V}^c\cap\mathcal{L}_{V|Y}^c,k}$ separately using a point-to-point linear non-capacity-achieving polar code with a uniform input distribution. The encoder here does not use the state information in the encoding process, but rather treat it as an unknown part of the channel noise.
\end{enumerate}

\noindent \textbf{Decoding}

\textbf{Input}: a noisy vector $y_{[n],[k]}\in\mathset{0,1}^kn$.

\textbf{Output:} a message estimation $\hat{m}_{[|\mathcal{H}_{V|S}\cap\mathcal{L}_{V|Y}|-|\mathcal{H}_{V|S}^c\cap\mathcal{L}_{V|Y}^c|],[k]}\in\mathset{0,1}^{k|\mathcal{H}_{V|S}\cap\mathcal{L}_{V|Y}|-|\mathcal{H}_{V|S}^c\cap\mathcal{L}_{V|Y}^c|}$.

\begin{enumerate}

\item Estimate the vector $u_{\mathcal{H}_{V|S}^c\cap\mathcal{L}_{V|Y}^c,k}$ by $\hat{u}_{\mathcal{H}_{V|S}^c\cap\mathcal{L}_{V|Y}^c,k}$, and let $\hat{u}_{\mathcal{R},k+1}=\hat{u}_{\mathcal{H}_{V|S}^c\cap\mathcal{L}_{V|Y}^c,k}$.

\item Estimate $u_{[n],[k]}$ by $\hat{u}_{[n],[k]}(y_{[n],[k]},f_{[|\mathcal{L}_{V|Y}^c\cap\mathcal{H}_{V|S}|],[k]})$ as follows: For each $j$ down from $k$ to $1$, and for each $i$ from 1 to $n$, assign
\begin{equation}
\label{eq:chaining_decoding}
\hat{u}_i^j= \left\{ 
  \begin{array}{l l}
    \displaystyle\argmax_{u\in\mathset{0,1}}p_{U_i|U_{[i-1]},Y_{[n]}}(u|u_{[i-1],j},y_{[n],j})& \quad \text{if $i\in \mathcal{L}_{V|Y}$}\\
		\hat{u}_{r(i,\mathcal{R}),j+1} &\quad \text{if $i\in \mathcal{H}_{V|S}^c\cap\mathcal{L}_{V|Y}^c$}\\
		f_{r(i,\mathcal{H}_{V|S}\cap\mathcal{L}_{V|Y}^c),j} & \quad \text{if $i\in \mathcal{H}_{V|S}\cap\mathcal{L}_{V|Y}^c$.}
  \end{array} \right.
\end{equation}
\item Return the estimated message $\hat{m}_{[|\mathcal{H}_{V|S}\cap\mathcal{L}_{V|Y}|-|\mathcal{H}_{V|S}^c\cap\mathcal{L}_{V|Y}^c|],[k]}=\hat{u}_{(\mathcal{H}_{V|S}\cap\mathcal{L}_{V|Y})\setminus\mathcal{R},[k]}$.
\end{enumerate}
\end{construction}

Constructions~\ref{con:multicoding} and~\ref{con:chaining} can also be used for communication over broadcast channels in Marton's region, as described in~\cite{GoeAbbGas13,MonHasSasUrb14}. Constructions~\ref{con:multicoding} and~\ref{con:chaining} improve on these previous results since they provably achieve the capacity with linear storage requirement.

Construction~\ref{con:chaining} achieves the capacity of Theorem~\ref{th:gp_capacity} with low complexity, without the degradation requirement of Property~\ref{pro:degradation}. This result was stated in Theorem~\ref{th:chaining}. The proof of Theorem~\ref{th:chaining} follows from Theorem~\ref{th:multicoding} and the fact that the rate loss vanishes with large $k$. Construction~\ref{con:chaining} is useful for the realistic model of flash memory-rewriting of Example~\ref{ex:realistic}, using the appropriate capacity-achieving functions $p(v|s)$ and $x(v,s)$. 

\section{Conclusion}
\label{sec:discussion}

In this paper we proposed three capacity-achieving polar coding schemes, for the settings of asymmetric channel coding and flash memory rewriting. The scheme for asymmetric channels improves on the scheme of~\cite{HonYam13} by reducing the exponential storage requirement into a linear one. The idea for this reduction is to perform the encoding randomly instead of using Boolean functions, and to transmit a vanishing fraction of information on a side channel. 

The second proposed scheme is used for the setting of flash memory rewriting. We propose a model of flash memory rewriting with writing noise, and show that the scheme achieves its capacity. We also describe a more general class of channels whose capacity can be achieved using the scheme. The second scheme is derived from the asymmetric-channel scheme by replacing the Shannon random variables $X$ and $X|Y$ with the Gelfand-Pinsker random variables $V|S$ and $V|Y$.

The last proposed scheme achieves the capacity of any channel with non-causal state information at the encoder. We bring a model of noisy flash memory rewriting for which the scheme would be useful. The main idea in this scheme is the code chaining proposed in~\cite{MonHasSasUrb14}. Another potential application could be in an asymmetric version of information embedding, which can be modeled as another special case of the Gelfand-Pinsker problem (as in~\cite{BarCheWor03}). 
%
%

\appendices

\section{}
\label{app:capacity}
In this appendix we provide an alternative proof for Theorem~\ref{th:degraded_capacity}, which was originally proven in \cite[Theorem 4]{Hee85}. We find the proof in this Appendix to be somewhat more intuitive. Theorem~\ref{th:degraded_capacity} states that the capacity of the channel in Example~\ref{ex:degraded} is 
$$C=(1-\beta)[h(\epsilon*\alpha)-h(\alpha)],$$
where $\epsilon=B/(1-\beta)$ and $\epsilon*\alpha\equiv \epsilon(1-\alpha)+(1-\epsilon)\alpha$.
An upper bound on the capacity can be obtained by assuming that the state information is available also to the decoder. In this case, the best coding scheme would ignore the cells with $s_i=1$ (about a fraction $\beta$ of the cells), and the rest of the cells would be coded according to a binary symmetric channel with an input cost constraint. It is optimal to assign a channel input $x_i=0$ for the cells with state $s_i=1$, such that those cells who do not convey information do not contribute to the cost. We now focus on the capacity of the binary symmetric channel with cost constraint. To comply with the expected input cost constraint $B$ of the channel of Example~\ref{ex:degraded}, the expected cost of the input to the binary symmetric channel (BSC) must be at most $\epsilon=B/(1-\beta)$. To complete the proof of the upper bound, we show next that the capacity of the BSC with cost constraint is equals to $h(\alpha\ast \epsilon)-h(\alpha)$. For this channel, we have 
$$H(Y|X)=h(\alpha)p_X(0)+h(\alpha)p_X(1)=h(\alpha).$$ 
We are left now with maximizing the entropy $H(Y)$ over the input pmfs $p_X(1)\le \epsilon$. We have
\begin{align*}
p_Y(1)=&p_{Y|X}(1|0)p_X(0)+p_{Y|X}(1|1)p_X(1)\\
=&\alpha(1-p_X(1))+(1-\alpha)p_X(1)\\
=&\alpha\ast p_X(1).
\end{align*}
Now since $p_X(1)\le \epsilon\le 1/2$ and $\alpha\ast p_X(1)$ is increasing in $p_X(1)$ below $1/2$, it follows that $p_Y(1)\le\alpha\ast \epsilon\le 1/2$ and therefore also that $H(Y)\le h(\alpha\ast \epsilon)$. So we have
$$\max_{p_X(1)\le\epsilon}I(X;Y)=\max_{p_X(1)\le\epsilon}(H(Y)-H(Y|X))=h(\alpha\ast \epsilon)-h(\alpha).$$
This completes the proof of the upper bound.

The lower bound is obtained by considering the selections $\mathcal{V}=\mathset{0,1}$, $x(v,0)=v$, $x(v,1)=0$ and
\begin{equation}
\label{eq:cond_dist_2}
p_{V|S}(1|0)=\epsilon, \qquad p_{V|S}(1|1)=\frac{\epsilon(1-\alpha)}{\epsilon*\alpha},
\end{equation}
and calculating the rate expression directly. Notice first that the cost constraint is met since
$$p_X(1)=p_{X|S}(1|0)p_S(0)=p_{V|S}(1|0)p_S(0)=\epsilon(1-\beta)=B.$$
We need to show that $H(V|S)-H(V|Y)=(1-\beta)[h(\alpha\ast \epsilon)- h(\alpha)]$.
Given the distributions $p_S$ and $p_{V|S}$, the conditional entropy
$H(V|S)$ is
\begin{align*}
H(V|S) =& \sum_{s\in\{0, 1\}}p_S(s) H(V|S = s)\notag\\
=& p_S(0) H(V|S = 0) + p_S(1) H(V|S = 1)\notag\\
=& (1-\beta) h(\epsilon) + \beta h\left(\frac{\epsilon(1-\alpha)}{\epsilon\ast \alpha}\right)
\end{align*}

To compute the conditional entropy $H(V|Y)$, we first compute the
probability distribution of the memory output $Y$ as follows:

\begin{align*}
p_Y(0) =& \sum_{v\in\{0, 1\}}p_{Y|VS}(0|v, 0)p_{V|S}(v |  0) p_S(0)\notag\\
=& (1-\beta) ((1-\alpha)(1-\epsilon) + \alpha \epsilon)\notag\\
=& (1-\beta)(\alpha\ast (1-\epsilon)),\\
p_Y(1) =& 1 - p_Y(0)\notag\\
=& (1-\beta)(\alpha\ast \epsilon) + \beta.
\end{align*}
The conditional distribution $p_{V|Y}$ is given by
\begin{align*}
p_{V|Y}(1 | 0)=& \sum_{s\in\{0, 1\}}p_{VS | Y}(1,s |0)\notag\\
=& \sum_{s\in\{0, 1\}}  \frac{p_{Y|VS}(0|1,s) p_{VS}(1,s) }{p_Y(0)}\notag\\ 
=& \sum_{s\in\{0, 1\}}  \frac{p_{Y|VS}(0|1,s) p_{V|S}(1 |s) p_S(s) }{p_Y(0)}\notag\notag\\
=& \frac{\alpha \epsilon}{\alpha\ast (1-\epsilon) },\\
p_{V|Y}(1 | 1)=& \sum_{s\in\{0, 1\}}p_{VS | Y}(1, s | 1)\notag\\
=& \sum_{s\in\{0, 1\}}  \frac{p_{Y|VS}(1|1, s) p_{VS}(1, s) }{p_Y(1)}\notag\\ 
=& \sum_{s\in\{0, 1\}}  \frac{p_{Y|VS}(1|1, s) p_{V|S}(1 | s) p_S( s) }{p_Y(1)}\notag\\
=& \frac{(1-\alpha)\epsilon(1-\beta) + \frac{\epsilon(1-\alpha)}{\epsilon\ast\alpha}\beta}{(1-\beta)(\alpha\ast \epsilon) + \beta}\notag\\
=& \frac{\epsilon(1-\alpha)}{\epsilon\ast\alpha}.
\end{align*}
Therefore we have
\begin{align*}
H(V|Y)=& \sum_{y\in\{0, 1\}}p_Y(y) H(V|Y = y)\notag\\
=& (1-\beta)(\alpha\ast (1-\epsilon))h\left(\frac{\alpha \epsilon}{\alpha\ast (1-\epsilon)}\right)+ (\beta + (1-\beta)(\alpha\ast \epsilon))h\left(\frac{\epsilon(1-\alpha)}{\epsilon\ast\alpha}\right),
\end{align*}
and then
\begin{align*}
H(V|S) - H(V|Y)=& (1-\beta) \bigg[h(\epsilon) - (\alpha\ast (1-\epsilon))h\left(\frac{\alpha \epsilon}{\alpha\ast (1-\epsilon)}\right)   - (\alpha\ast \epsilon)h\left(\frac{\epsilon(1-\alpha)}{\epsilon\ast\alpha}\right)\bigg]\notag\\
=& (1-\beta)\bigg[h(\epsilon) + \alpha \epsilon\log_2 \frac{\alpha \epsilon}{\alpha\ast(1-\epsilon)} + (1-\alpha)(1-\epsilon)\log_2 \frac{(1-\alpha)(1-\epsilon)}{\alpha\ast(1-\epsilon)}\notag\\
&\ \ \ \ \ \ \ \ + \alpha(1-\epsilon)\log_2 \frac{\alpha(1-\epsilon)}{\alpha\ast \epsilon} + \epsilon(1- \alpha)\log_2 \frac{\epsilon(1-\alpha)}{\alpha\ast \epsilon}\bigg]\notag\\
=& (1 - \beta)[ h(\alpha\ast \epsilon) + h(\epsilon) +\alpha \epsilon\log_2(\alpha \epsilon) + (1-\alpha)(1-\epsilon)\log_2(1-\alpha)(1-\epsilon)\notag\\ 
&\ \ \ \ \ \ \ \ \ \ + \alpha(1-\epsilon)\log_2\alpha(1-\epsilon)+ \epsilon(1-\alpha)\log_2\epsilon(1-\alpha)]\notag\\
=& (1-\beta)\left[h(\alpha\ast \epsilon) + h(\epsilon) - h(\alpha)  - h(\epsilon)\right]\notag\\
=& (1-\beta)\left[h(\alpha\ast \epsilon) - h(\alpha)\right].
\end{align*}

\section{}
\label{app:degradation}

In this appendix we prove Lemma~\ref{lem:degradation}. We need to show that, using the functions of Theorem~\ref{th:degraded_capacity}, there exists a DMC $W:\mathset{0,1}\to\mathset{0,1}$ such that 
\begin{equation}
\label{eq:degradation}
p_{{S}|{V}}({s}|{v})=\sum_{{y}\in\mathset{0,1}}p_{{Y}|{V}}({y}|{v})W({s}|{y}).
\end{equation}
To define such channel $W$, we first claim that
\begin{equation}
\label{eq:relation}
p_{Y|V,S}(1|v,0)p_{V|S}(v|0)=(\epsilon*\alpha)p_{V|S}(v|1).
\end{equation}
Equation (\ref{eq:relation}) follows directly from Equation (\ref{eq:cond_dist}) since
\begin{equation*}
\frac{p_{Y|V,S}(1|0,0)p_{V|S}(0|0)}{p_{V|S}(0|1)}=\frac{\alpha(1-\epsilon)}{\frac{\alpha(1-\epsilon)}{\epsilon*\alpha}}=\epsilon*\alpha,
\end{equation*}
\begin{equation*}
\frac{p_{Y|V,S}(1|1,0)p_{V|S}(1|0)}{p_{V|S}(1|1)}=\frac{(1-\alpha)\epsilon}{\frac{(1-\alpha)\epsilon}{\epsilon*\alpha}}=\epsilon*\alpha.
\end{equation*}

Next, we claim that $\frac{p_{V,S}(v,1)}{p_{V,Y}(v,1)}=\frac{\beta}{(\epsilon*\alpha)(1-\beta)+\beta}$ for any $v\in\mathset{0,1}$, and therefore that $\frac{p_{V,S}(v,1)}{p_{V,Y}(v,1)}\in[0,1]$. This follows from
\begin{align*}
\frac{p_{V,S}(v,1)}{p_{V,Y}(v,1)}\overset{\text{(a)}}=&\frac{p_{V|S}(v|1)p_S(1)}{p_{Y,V|S}(1,v|0)p_S(0)+p_{Y,V|S}(1,v|1)p_S(1)}\\
\overset{\text{(b)}}=&\frac{p_{V|S}(v|1)\beta}{p_{Y|V,S}(1|v,0)p_{V|S}(v|0)(1-\beta)+p_{Y|V,S}(1|v,1)p_{V|S}(x|1)\beta}\\
\overset{\text{(c)}}=&\frac{p_{V|S}(v|1)\beta}{(\epsilon*\alpha)p_{V|S}(v|1)(1-\beta)+p_{V|S}(v|1)\beta}\\
=&\frac{\beta}{(\epsilon*\alpha)(1-\beta)+\beta},
\end{align*}
where (a) follows from the law of total probability, (b) follows from the definition of conditional probability, and (c) follows from Equations (\ref{eq:channel_given_state}) and (\ref{eq:relation}).

Since $\frac{p_{V,S}(v,1)}{p_{V,Y}(v,1)}$
is not a function of $v$ and is in $[0,1]$, we can define $W$ as
following:
\begin{equation*}
W({s}|{y})\triangleq\left\{ 
  \begin{array}{l l}
    1 & \quad \text{if  $(s,y)=(0,0)$}\\
    1-\frac{p_{S|V}(1|v)}{p_{Y|V}(1|v)} & \quad \text{if  $(s,y)=(0,1)$}\\
		\frac{p_{S|V}(1|v)}{p_{Y|V}(1|v)} & \quad \text{if  $(s,y)=(1,1)$}\\
		0 & \quad \text{if  $(s,y)=(1,0)$.}
  \end{array} \right.
\end{equation*}
We show next that Equation~(\ref{eq:degradation}) holds for $W$ defined above:
\begin{align*}
\sum_{{y}\in\mathset{0,1}}p_{Y|V}(y|{v})W({s}|{y})=&p_{Y|V}(0|v)W(s|0)+p_{Y|V}(1|v)W(s|1)\\
=&\left[p_{Y|V}(0|v)+p_{Y|V}(1|v)\left(1-\frac{p_{S|V}(1|v)}{p_{Y|V}(1|v)}\right)\right]\mathbb{1}[s=0]+p_{S|V}(1|v)\mathbb{1}[s=1]\\
=&\left[1-p_{S|V}(1|v)\right]\mathbb{1}[s=0]+p_{S|V}(1|v)\mathbb{1}[s=1]\\
=&p_{S|V}(0|v)\mathbb{1}[s=0]+p_{S|V}(1|v)\mathbb{1}[s=1]\\
=&p_{S|V}(s|v).
\end{align*}
 So the channel $W$ satisfies Equation (\ref{eq:degradation}) and thus the lemma holds.

\section{}
\label{app:multicoding}

In this appendix we prove Theorem~\ref{th:multicoding}. The complexity claim of Theorem \ref{th:multicoding} is explained in \cite[Section III.B]{HonYam13}. We start with the asymptotic rate of Construction~\ref{con:multicoding}. We want to show that $\lim_{n\to\infty}(1/n)|\mathcal{H}_{V|S}^c\cap\mathcal{L}_{V|Y}^c|=0$. Since $p_{S|V}$ is degraded with respect to $p_{Y|V}$, it follows from~\cite[Lemma 4]{GoeAbbGas13} that $\mathcal{L}_{V|S}\subseteq\mathcal{L}_{V|Y}$, and therefore that $\mathcal{L}_{V|S}^c\supseteq\mathcal{L}_{V|Y}^c$. So we have 
$$\lim_{n\to\infty}(1/n)|\mathcal{H}_{V|S}^c\cap\mathcal{L}_{V|Y}^c|\le\lim_{n\to\infty}(1/n)|\mathcal{H}_{V|S}^c\cap\mathcal{L}_{V|S}^c|=0,$$
where the equality follows from the definition of the sets. To complete the proof of the theorem, we need to show that for some frozen vector, the input cost meets the design constraint, while the decoding error probability vanishes sub-exponentially fast in the block length. We show this result using the probabilistic method. That is, we first assume that the frozen vector is random, drawn uniformly, and analyze the input cost and error probability in this case, and then we show the existence of a good vector using Markov's inequalities and the union bound. Denote the uniformly distributed random vector that represents the frozen vector by $F_{[|\mathcal{H}_{V|S}\cap\mathcal{L}_{V|Y}^c|]}$. We show next that the expected input cost exceeds the design constraint $B$ by a vanishing amount.


\subsection{Expected Input Cost}

Define 
$$b^n(X_{[n]})=\sum_{i=1}^n b(X_i).$$
For a state vector $s_{[n]}$ and the encoding rule (\ref{eq:encoding}) with uniform frozen vector $F_{[|\mathcal{H}_{V|S}\cap\mathcal{L}_{V|Y}^c|]}$, each vector $u_{[n]}$ appears with probability
\begin{equation*}
\left(\prod_{i\in\mathcal{H}_{V|S}^c}p_{U_i|U_{[i-1]},S_{[n]}}(u_i|u_{[i-1]},s_{[n]})\right)2^{-|\mathcal{H}_{V|S}|}.
\end{equation*}
Remember that the joint pmf $p_{S_{[n]},U_{[n]}}$ refers to the capacity-achieving distribution of the channel. The expected cost is expressed as 
%
%

%
\begin{align*}
\mathsf{E}_{F_{[|\mathcal{H}_{V|S}\cap\mathcal{L}_{V|Y}^c|]}}(b^n(X_{[n]}))=&\sum_{u_{[n]},s_{[n]}}p_{S_{[n]}}(s_{[n]})\left(\prod_{i\in\mathcal{H}_{V|S}^c}p_{U_i|U_{[i-1]},S_{[n]}}(u_i|u_{[i-1]},s_{[n]})\right)2^{-|\mathcal{H}_{V|S}|} b^n(u_{[n]}G_n).
\end{align*}
Define the joint pmf
\begin{align}
\label{eq:q}
q_{S_{[n]},U_{[n]}}\equiv p_{S_{[n]}}(s_{[n]})\left(\prod_{i\in\mathcal{H}_{V|S}^c}p_{U_i|U_{[i-1]},S_{[n]}}(u_i|u_{[i-1]},s_{[n]})\right)2^{-|\mathcal{H}_{V|S}|}.
\end{align}
Intuitively, $q_{U_{[n]},S_{[n]}}$ refers to the distribution imposed by the encoding procedure of Construction~\ref{con:multicoding}. Then we have
\begin{align*}
\mathsf{E}_{F_{[|\mathcal{H}_{V|S}\cap\mathcal{L}_{V|Y}^c|]}}(b^n(X_{[n]}))
=&\mathsf{E}_{q_{U_{[n]},S_{[n]}}}[b^n(U_{[n]}G_n)]\\
\le& \mathsf{E}_{p_{U_{[n]},S_{[n]}}}[b^n(U_{[n]}G_n)]+\max_x b(x)\|p_{U_{[n]},S_{[n]}}-q_{U_{[n]},S_{[n]}}\|\\
=& nB+\max_x b(x)\|p_{U_{[n]},S_{[n]}}-q_{U_{[n]},S_{[n]}}\|,
\end{align*}
where $\|\cdot\|$ is the $L_1$ distance and the inequality follows from the triangle inequality. We will now prove that 
$$\mathsf{E}_{F_{[|\mathcal{H}_{V|S}\cap\mathcal{L}_{V|Y}^c|]}}(b^n(X_{[n]}))\le nB +2^{-n^{1/2-\delta}},$$ by showing that $\|p_{U_{[n]},S_{[n]}}-q_{U_{[n]},S_{[n]}}\|\le 2^{-n^{1/2-\delta}}$. To do this, we will prove a slightly stronger relation that will be used also for the proof of the probability of decoding error. We first define the joint pmf
\begin{align*}
q_{U_{[n]},S_{[n]},Y_{[n]}}\equiv q_{U_{[n]},S_{[n]}}(u_{[n]},s_{[n]}) p_{Y_{[n]}|U_{[n]},S_{[n]}}(y_{[n]}|u_{[n]},s_{[n]}).
\end{align*}
Then we notice that 
\begin{align*}
\|p_{U_{[n]},S_{[n]}}-q_{U_{[n]},S_{[n]}}\|=&\sum_{u_{[n]},s_{[n]}}|p(u_{[n]},s_{[n]})-q(u_{[n]},s_{[n]})|\notag\\
=&\sum_{u_{[n]},s_{[n]}}\left|\sum_{y_{[n]}}[p(u_{[n]},s_{[n]},y_{[n]})-q(u_{[n]},s_{[n]},y_{[n]})]\right|\notag\\
\le&\sum_{u_{[n]},s_{[n]},y_{[n]}}|p(u_{[n]},s_{[n]},y_{[n]})-q(u_{[n]},s_{[n]},y_{[n]})|\notag\\
=&\|p_{U_{[n]},S_{[n]},Y_{[n]}}-q_{U_{[n]},S_{[n]},Y_{[n]}}\|,
\end{align*}
where the inequality follows from the triangle inequality. The proof of the expected cost is completed with the following lemma, which will be used also for analyzing the probability of decoding error.

\begin{lemma}
\label{lem:distance}
\begin{equation}
\|p_{U_{[n]},S_{[n]},Y_{[n]}}-q_{U_{[n]},S_{[n]},Y_{[n]}}\|\le 2^{-n^{1/2-\delta}}.
\end{equation}
\end{lemma}
\begin{IEEEproof}
Let $D(\cdot\|\cdot)$ denote the relative entropy. Then
\begin{align}
\label{eq:distance}
\|p_{U_{[n]},S_{[n]},Y_{[n]}}&-q_{U_{[n]},S_{[n]},Y_{[n]}}\|\notag\\
=&\sum_{u_{[n]},s_{[n]},y_{[n]}}|p(u_{[n]},s_{[n]},y_{[n]})-q(u_{[n]},s_{[n]},y_{[n]})|\notag\\
\overset{\text{(a)}}=&\sum_{u_{[n]},s_{[n]},y_{[n]}}|p(u_{[n]}|s_{[n]})-q(u_{[n]}|s_{[n]})|p(s_{[n]})p(y_{[n]}|u_{[n]},s_{[n]})\notag\\
\overset{\text{(b)}}=&\sum_{u_{[n]},s_{[n]},y_{[n]}}\left|\prod_{i=1}^{n} p(u_i|u_{[i-1]},s_{[n]})-\prod_{i=1}^{n} q(u_i|u_{[i-1]},s_{[n]})
\right| p(s_{[n]})p(y_{[n]}|u_{[n]},s_{[n]})\notag\\
\overset{\text{(c)}}=&\sum_{u_{[n]},s_{[n]},y_{[n]}}\sum_{i=1}^{n} \left|[p(u_i|u_{[i-1]},s_{[n]})- q(u_i|u_{[i-1]},s_{[n]})]\right|p(s_{[n]})p(y_{[n]}|u_{[n]},s_{[n]})\notag\\
&\cdot\prod_{j=1}^{i-1} p(u_j|u_{[j-1]},s_{[n]})\prod_{j=i+1}^{n} q(u_j|u_{[j-1]},s_{[n]}) \notag\\
\overset{\text{(d)}}\le&\sum_{i\in\mathcal{H}_{V|S}}\sum_{u_{[n]},s_{[n]},y_{[n]}}\left| p(u_i|u_{[i-1]},s_{[n]})- q(u_i|u_{[i-1]},s_{[n]})\right|p(s_{[n]})\prod_{j=1}^{i-1} p(u_j|u_{[j-1]},s_{[n]})\notag\\
&\cdot\prod_{j=i+1}^{n} q(u_j|u_{[j-1]},s_{[n]}) p(y_{[n]}|u_{[n]},s_{[n]})\notag\\
=&\sum_{i\in\mathcal{H}_{V|S}}\sum_{u_{[i]},s_{[n]}}\left| p(u_i|u_{[i-1]},s_{[n]})- q(u_i|u_{[i-1]},s_{[n]})\right|\prod_{j=1}^{i-1} p(u_j|u_{[j-1]},s_{[n]}) p(s_{[n]})\notag\\
\overset{\text{(e)}}=&\sum_{i\in\mathcal{H}_{V|S}}\sum_{u_{[i-1]},s_{[n]}}p(u_{[i-1]},s_{[n]}) \|p_{U_i|U_{[i-1]}=u_{[i-1]},S_{[n]}=s_{[n]}}-q_{U_i|U_{[i-1]}=u_{[i-1]},S_{[n]}=s_{[n]}}\|\notag\\
\overset{\text{(f)}}\le&\sum_{i\in\mathcal{H}_{V|S}}\sum_{u_{[i-1]},s_{[n]}}p(u_{[i-1]},s_{[n]})\sqrt{2\ln 2}\sqrt{D(p_{U_i|U_{[i-1]}=u_{[i-1]},S_{[n]}=s_{[n]}}\|q_{U_i|U_{[i-1]}=u_{[i-1]},S_{[n]}=s_{[n]}})}\notag\\
\overset{\text{(g)}}\le&\sum_{i\in\mathcal{H}_{V|S}} \sqrt{(2\ln 2)\sum_{u_{[i-1]},s_{[n]}}p(u_{[i-1]},s_{[n]}) D(p_{U_i|U_{[i-1]}=u_{[i-1]},S_{[n]}=s_{[n]}}\|q_{U_i|U_{[i-1]}=u_{[i-1]},S_{[n]}=s_{[n]}})
}\notag\\
=&\sum_{i\in\mathcal{H}_{V|S}}\sqrt{(2\ln 2) D(p_{U_i}\|q_{U_i}|U_{[i-1]},S_{[n]})}\notag\\
\overset{\text{(h)}}=&\sum_{i\in \mathcal{H}_{V|S}}\sqrt{(2\ln 2) [1-H(U_i|U_{[i-1]},S_{[n]})]}\notag,
\end{align} 
where
\begin{enumerate}[(a)]
\item follows from the fact that $p(s_{[n]})=q(s_{[n]})$ and $p(y_{[n]}|u_{[n]},s_{[n]})=q(y_{[n]}|u_{[n]},s_{[n]})$,
\item follows from the chain rule,
\item follows from the telescoping expansion
\begin{align*}
B_{[n]}-A_{[n]}&=\sum_{i=1}^n A_{[i-1]}B_{[i:n]}-\sum_{i=1}^n A_{[i]}B_{[i+1:n]}\notag\\
&=\sum_{i=1}^n (B_i-A_i)A_{[i-1]}B_{[i+1:n]},
\end{align*}
where $A_{[j:k]}$ and $B_{[j:k]}$ denote the products $\prod_{i=j}^k A_i$ and $\prod_{i=j}^k B_i$, respectively,
\item follows from the triangular inequality and the fact that $p(u_i|u_{[i-1]},s_{[n]})=q(u_i|u_{[i-1]},s_{[n]})$ for all $i\in\mathcal{H}_{V|S}^c$ (as defined in Equation (\ref{eq:q})),
\item follows from the chain rule again,
\item follows from Pinsker's inequality (see, e.g., \cite[Lemma 11.6.1]{CovTho06}),
\item follows from Jensen's inequality and
\item follows from the facts that $q(u_i|u_{[i-1]},s_{[n]})=1/2$ for $i\in \mathcal{H}_{V|S}$ and from~\cite[Lemma 10]{GoeAbbGas13}.
\end{enumerate}

Now if $i\in \mathcal{H}_{V|S}$, we have
\begin{align}
1-H(U_i|U_{[i-1]},S_{[n]})&\le 1-[Z(U_i|U_{[i-1]},S_{[n]})]^2\notag\\
&\le  2^{-2n^{1/2-\delta}},
\end{align}
where the first inequality follows from Proposition \ref{prop:entropy}, and the second inequality follows from the fact that $i$ is in $\mathcal{H}_{V|S}$. This completes the proof of the lemma.
\end{IEEEproof}


\subsection{Probability of Decoding Error}
\label{sub:prob}


Let $\mathcal{E}_i$ be the set of pairs of vectors $(u_{[n]},y_{[n]})$ such that $\hat{u}_{[n]}$ is a result of decoding $y_{[n]}$, and $\hat{u}_{[i]}$ satisfies both $\hat{u}_{[i-1]}= u_{[i-1]}$  and $\hat{u}_i\ne u_i$. The block decoding error event is given by $\mathcal{E}\equiv \cup_{i\in \mathcal{L}_{V|Y}}\mathcal{E}_i$. Under decoding given in (\ref{eq:multidecoding}) with an arbitrary tie-breaking rule, every pair $(u_{[n]},y_{[n]})\in\mathcal{E}_i$ satisfies
\begin{equation}
p_{U_i|U_{[i-1]},Y_{[n]}}(u_i|u_{[i-1]},y_{[n]})\le p_{U_i|U_{[i-1]},Y_{[n]}}(u_i\oplus 1|u_{[i-1]},y_{[n]}).
\end{equation}

Consider the block decoding error probability $p_e(F_{[|\mathcal{H}_{V|S}\cap\mathcal{L}_{V|Y}^c|]})$ for the random frozen vector $F_{[|\mathcal{H}_{V|S}\cap\mathcal{L}_{V|Y}^c|]}$. For a state vector $s_{[n]}$ and the encoding rule (\ref{eq:encoding}), each vector $u_{[n]}$ appears with probability
\begin{equation*}
\left(\prod_{i\in\mathcal{H}_{V|S}^c}p_{U_i|U_{[i-1]},S_{[n]}}(u_i|u_{[i-1]},s_{[n]})\right)2^{-|\mathcal{H}_{V|S}|}.
\end{equation*}
By the definition of conditional probability and the law of total probability, the probability of error $p_e(F_{[|\mathcal{H}_{V|S}\cap\mathcal{L}_{V|Y}^c|]})$ is given by
%
\begin{align*}
\mathsf{E}_{F_{[|\mathcal{H}_{V|S}\cap\mathcal{L}_{V|Y}^c|]}}[p_e]
=\sum_{u_{[n]},s_{[n]},y_{[n]}}&p_{S_{[n]}}(s_{[n]})\left(\prod_{i\in\mathcal{H}_{V|S}^c}p_{U_i|U_{[i-1]},S_{[n]}}(u_i|u_{[i-1]},s_{[n]})\right)2^{-|\mathcal{H}_{V|S}|}\\
&\cdot p_{Y_{[n]}|U_{[n]},S_{[n]}}(y_{[n]}|u_{[n]},s_{[n]})\mathbb{1}[(u_{[n]},y_{[n]})\in\mathcal{E}].
\end{align*}
Then we have
\begin{align*}
\mathsf{E}_{F_{[|\mathcal{H}_{V|S}\cap\mathcal{L}_{V|Y}^c|]}}[p_e]
=&q_{U_{[n]},Y_{[n]}}(\mathcal{E})\\
\le&\|q_{U_{[n]},Y_{[n]}}-p_{U_{[n]},Y_{[n]}}\|+p_{U_{[n]},Y_{[n]}}(\mathcal{E})\notag\\
\le&\|q_{U_{[n]},Y_{[n]}}-p_{U_{[n]},Y_{[n]}}\|+\sum_{i\in \mathcal{L}_{V|Y}}p_{U_{[n]},Y_{[n]}}(\mathcal{E}_i),
\end{align*}
where the first inequality follows from the triangle inequality. Each term in the summation is bounded by
\begin{align*}
p_{U_{[n]},Y_{[n]}}(\mathcal{E}_i)&\le\sum_{u_{[i]},y_{[n]}}p(u_{[i]},y_{[n]})\mathbb{1}\left[p(u_i|u_{[i-1]},y_{[n]})\le p(u_i\oplus 1|u_{[i-1]},y_{[n]})\right]\notag\\
&\le\sum_{u_{[i]},y_{[n]}}p(u_{[i-1]},y_{[n]})p(u_i|u_{[i-1]},y_{[n]})
\sqrt{\frac{p(u_i\oplus 1|u_{[i-1]},y_{[n]})}{p(u_i|u_{[i-1]},y_{[n]})}}\notag\\
&=Z(U_i|U_{[i-1]},Y_{[n]})\notag\\
&\le 2^{-n^{1/2-\delta}},
\end{align*}
where the last inequality follows from the fact that $i$ belongs to the set $\mathcal{L}_{V|Y}$. 

To prove that $\mathsf{E}_{F_{[|\mathcal{H}_{V|S}\cap\mathcal{L}_{V|Y}^c|]}}[p_e]\le 2^{-n^{1/2-\delta'}}$ for some $\delta'>\delta$, we are left with showing that $\|p_{U_{[n]},Y_{[n]}}-q_{U_{[n]},Y_{[n]}}\|\le 2^{-n^{1/2-\delta}}$. Notice that 
\begin{align*}
2\|p_{U_{[n]},Y_{[n]}}-q_{U_{[n]},Y_{[n]}}\|=&\sum_{u_{[n]},y_{[n]}}|p(u_{[n]},y_{[n]})-q(u_{[n]},y_{[n]})|\notag\\
=&\sum_{u_{[n]}, y_{[n]}}\left|\sum_{s_{[n]}}[p(u_{[n]},s_{[n]},y_{[n]})-q(u_{[n]},s_{[n]},y_{[n]})]\right|\notag\\
\le&\sum_{u_{[n]},s_{[n]},y_{[n]}}|p(u_{[n]},s_{[n]},y_{[n]})-q(u_{[n]},s_{[n]},y_{[n]})|\notag\\
=&2\|p_{U_{[n]},S_{[n]},Y_{[n]}}-q_{U_{[n]},S_{[n]},Y_{[n]}}\|,
\end{align*}
where the inequality follows from the triangle inequality. Lemma~\ref{lem:distance} now completes the proof that $\mathsf{E}_{F_{[|\mathcal{H}_{V|S}\cap\mathcal{L}_{V|Y}^c|]}}[p_e]=2^{-n^{1/2-\delta}}$. 

\subsection{Existence of a Good Frozen Vector}
\label{app:existance}

We showed that
$$\mathsf{E}_{F_{[|\mathcal{H}_{V|S}\cap\mathcal{L}_{V|Y}^c|]}}(b^n(X_{[n]}))\le nB +2^{-n^{1/2-\delta}}.$$
In this subsection we will be interested to find a frozen vector which satisfies a slightly higher expected cost, namely
$$\mathsf{E}(b^n(X_{[n]}))\le nB + 1/n^2.$$
Remember that $B$ is a constant. We take a uniformly distributed frozen vector, and bound the probability that the expected cost exceeds $nB + 1/n^2$, using Markov's inequality. The following inequalities hold for large enough $n$, and we are not concerned here with the exact values required.
\begin{align*}
P(\mathsf{E}(b^n(X_{[n]}))\ge nB + 1/n^2) &\le \mathsf{E}_{F_{[|\mathcal{H}_{V|S}\cap\mathcal{L}_{V|Y}^c|]}}(b^n(X))/(nB + 1/n^2)\\
&\le (nB +2^{-n^{1/2-\delta}})/(nB + 1/n^2)\\
&\le (nB +n^{-3})/(nB + n^{-2})\\
&= 1-(n^{-2}-n^{-3})/(nB + n^{-2})\\
&\le 1-(n^{-2.1})/(nB + n^{-2})\\
&\le 1-1/(Bn^{3.1} + n^{0.1})\\
&\le 1-1/(Bn^{3.2} )\\
&\le 1-1/n^{4}.
\end{align*}
We now apply Markov's inequality on the probability of decoding error:
$$P(p_e\ge n^5 2^{-n^{1/2-\delta}})\le \frac{E_{F_{[|\mathcal{H}_{V|S}\cap\mathcal{L}_{V|Y}^c|]}}(p_e)}{n^5 2^{-n^{1/2-\delta}}}\le\frac{2^{-n^{1/2-\delta}}}{n^5 2^{-n^{1/2-\delta}}} = 1/n^5.$$
By the union bound, the probability that either $\mathsf{E}(b^n(X_{[n]}))\ge nB + 1/n^2$ or $p_e\ge n^5 2^{-n^{1/2-\delta}}$ is at most 
$$1-1/n^4+1/n^5\le 1-1/n^5.$$
This implies that the probability that both $\mathsf{E}(b^n(X_{[n]}))\le nB + 1/n^2$ and $p_e\le n^5 2^{-n^{1/2-\delta}}$ is at least $1/n^5$. So there exists at least one frozen vector for which the desired properties for both the expected cost and the probability of error hold. Furthermore, such a vector can be found by repeatedly selecting a frozen vector uniformly at random, until the required properties hold. The properties can be verified efficiently with close approximations by the upgradation and degradation method proposed in~\cite{TalVardy13}. The expected number of times until a good vector is found is polynomial in the block length (at most $n^5$).  This completes the proof of Theorem~\ref{th:multicoding}.

%

\bibliographystyle{IEEEtranS}
\bibliography{IEEEabrv,local}

\begin{thebibliography}{10}
\providecommand{\url}[1]{#1}
\csname url@samestyle\endcsname
\providecommand{\newblock}{\relax}
\providecommand{\bibinfo}[2]{#2}
\providecommand{\BIBentrySTDinterwordspacing}{\spaceskip=0pt\relax}
\providecommand{\BIBentryALTinterwordstretchfactor}{4}
\providecommand{\BIBentryALTinterwordspacing}{\spaceskip=\fontdimen2\font plus
\BIBentryALTinterwordstretchfactor\fontdimen3\font minus
  \fontdimen4\font\relax}
\providecommand{\BIBforeignlanguage}[2]{{%
\expandafter\ifx\csname l@#1\endcsname\relax
\typeout{** WARNING: IEEEtranS.bst: No hyphenation pattern has been}%
\typeout{** loaded for the language `#1'. Using the pattern for}%
\typeout{** the default language instead.}%
\else
\language=\csname l@#1\endcsname
\fi
#2}}
\providecommand{\BIBdecl}{\relax}
\BIBdecl

\bibitem{Ari09}
E.~Arikan, ``Channel polarization: A method for constructing capacity-achieving
  codes for symmetric binary-input memoryless channels,'' \emph{IEEE Trans.
  Inf. Theory}, vol.~55, no.~7, pp. 3051--3073, July 2009.

\bibitem{Ari10}
------, ``Source polarization,'' in \emph{Proc. IEEE Int. Symp. on Inf. Theor.
  (ISIT)}, June 2010, pp. 899--903.

\bibitem{BarCheWor03}
R.~Barron, B.~Chen, and G.~W. Wornell, ``The duality between information
  embedding and source coding with side information and some applications,''
  \emph{IEEE Trans. Inf. Theory}, vol.~49, no.~5, pp. 1159--1180, May 2003.

\bibitem{BurStr13}
D.~Burshtein and A.~Strugatski, ``Polar write once memory codes,'' \emph{IEEE
  Trans. Inf. Theory}, vol.~59, no.~8, pp. 5088--5101, Aug. 2013.

\bibitem{CasSchBohBru10}
Y.~Cassuto, M.~Schwartz, V.~Bohossian, and J.~Bruck, ``Codes for asymmetric
  limited-magnitude errors with application to multilevel flash memories,''
  \emph{IEEE Trans. Inf. Theory}, vol.~56, no.~4, pp. 1582--1595, April 2010.

\bibitem{CovTho06}
T.~M. Cover and J.~A. Thomas, \emph{Elements of information theory (2.
  ed.)}.\hskip 1em plus 0.5em minus 0.4em\relax Wiley, 2006.

\bibitem{ElgKim12}
A.~{El Gamal} and Y.~Kim, \emph{Network information theory}.\hskip 1em plus
  0.5em minus 0.4em\relax Cambridge University Press, 2012.

\bibitem{EngLiKliLanJiaBru14a}
E.~En~Gad, Y.~Li, J.~Kliewer, M.~Langberg, A.~Jiang, and J.~Bruck, ``Polar
  coding for noisy write-once memories,'' in \emph{Proc. IEEE Int. Symp. on
  Inf. Theor. (ISIT)}, June 2014, pp. 1638--1642.

\bibitem{EngYaaJiaBru13}
E.~En~Gad, E.~Yaakobi, A.~Jiang, and J.~Bruck, ``Rank-modulation rewrite coding
  for flash memories,'' \emph{Available at http://arxiv.org/abs/1312.0972}, Dec
  2013.

\bibitem{GabSha12}
A.~Gabizon and R.~Shaltiel, ``Invertible zero-error dispersers and defective
  memory with stuck-at errors,'' in \emph{APPROX-RANDOM}, 2012, pp. 553--564.

\bibitem{Gal68}
R.~G. Gallager, \emph{Information Theory and Reliable Communication}.\hskip 1em
  plus 0.5em minus 0.4em\relax Wiley, 1968.

\bibitem{GelPin80}
S.~Gel'fand and M.~Pinsker, ``Coding for channel with random parameters,''
  \emph{Problems of Control Theory}, vol.~9, no.~1, pp. 19--31, 1980.

\bibitem{GoeAbbGas13}
N.~Goela, E.~Abbe, and M.~Gastpar, ``Polar codes for broadcast channels,''
  \emph{Available at http://arxiv.org/abs/1301.6150}, Jan 2013.

\bibitem{Gol80}
S.~Golomb, ``The limiting behavior of the z-channel (corresp.),'' \emph{IEEE
  Trans. Inf. Theory}, vol.~26, no.~3, pp. 372--372, May 1980.

\bibitem{Gor62}
J.~Gordon, ``Quantum effects in communications systems,'' \emph{Proceedings of
  the IRE}, vol.~50, no.~9, pp. 1898--1908, Sept 1962.

\bibitem{Hee85}
C.~Heegard, ``On the capacity of permanent memory,'' \emph{IEEE Trans. Inf.
  Theory}, vol.~31, no.~1, pp. 34--42, January 1985.

\bibitem{HonYam13}
J.~Honda and H.~Yamamoto, ``Polar coding without alphabet extension for
  asymmetric models,'' \emph{IEEE Trans. Inf. Theory}, vol.~59, no.~12, pp.
  7829--7838, 2013.

\bibitem{KorUrb10}
S.~Korada and R.~Urbanke, ``Polar codes are optimal for lossy source coding,''
  \emph{IEEE Trans. Inf. Theory}, vol.~56, no.~4, pp. 1751--1768, April 2010.

\bibitem{KusTsy74}
A.~Kusnetsov and B.~S. Tsybakov, ``Coding in a memory with defective cells,,''
  \emph{translated from Problemy Peredachi Informatsii}, vol.~10, no.~2, pp.
  52--60, 1974.

\bibitem{Ma14}
X.~Ma, ``Write-once-memory codes by source polarization,'' \emph{Available at
  http://arxiv.org/abs/1405.6262}, May 2014.

\bibitem{Mar79}
K.~Marton, ``A coding theorem for the discrete memoryless broadcast channel,''
  \emph{IEEE Trans. Inf. Theory}, vol.~25, no.~3, pp. 306--311, May 1979.

\bibitem{MonHasUrb14}
M.~Mondelli, I.~Hassani, and R.~Urbanke, ``How to achieve the capacity of
  asymmetric channels,'' \emph{Available at http://arxiv.org/abs/1406.7373},
  June 2014.

\bibitem{MonHasSasUrb14}
M.~Mondelli, S.~H. Hassani, I.~Sason, and R.~Urbanke, ``Achieving {Marton's}
  region for broadcast channels using polar codes,'' \emph{Available at
  http://arxiv.org/abs/1401.6060}, January 2014.

\bibitem{MorTan10}
R.~Mori and T.~Tanaka, ``Channel polarization on q-ary discrete memoryless
  channels by arbitrary kernels,'' in \emph{IEEE Int. Symp. on Inf. Theor.
  (ISIT)}, June 2010, pp. 894--898.

\bibitem{ParBar13}
W.~Park and A.~Barg, ``Polar codes for q-ary channels, $q=2^{r}$,'' \emph{IEEE
  Trans. Inf. Theory}, vol.~59, no.~2, pp. 955--969, Feb 2013.

\bibitem{SahPra11}
A.~Sahebi and S.~Pradhan, ``Multilevel polarization of polar codes over
  arbitrary discrete memoryless channels,'' in \emph{Communication, Control,
  and Computing (Allerton), 2011 49th Annual Allerton Conference on}, Sept
  2011, pp. 1718--1725.

\bibitem{Sas12a}
E.~Sasoglu, ``Polar codes for discrete alphabets,'' in \emph{Proc. IEEE Int.
  Symp. on Inf. Theor. (ISIT)}, July 2012, pp. 2137--2141.

\bibitem{SasTelAri09}
E.~Sasoglu, I.~Telatar, and E.~Arikan, ``Polarization for arbitrary discrete
  memoryless channels,'' in \emph{IEEE Information Theory Workshop (ITW)}, Oct
  2009, pp. 144--148.

\bibitem{Sas12}
\BIBentryALTinterwordspacing
E.~Sasoglu, ``Polarization and polar codes,'' \emph{Foundations and Trends in
  Communications and Information Theory}, vol.~8, no.~4, pp. 259--381, 2012.
  [Online]. Available: \url{http://dx.doi.org/10.1561/0100000041}
\BIBentrySTDinterwordspacing

\bibitem{SutRenDupRen12}
D.~Sutter, J.~Renes, F.~Dupuis, and R.~Renner, ``Achieving the capacity of any
  {DMC} using only polar codes,'' in \emph{IEEE Information Theory Workshop
  (ITW)}, Sept 2012, pp. 114--118.

\bibitem{TalVardy13}
I.~Tal and A.~Vardy, ``How to construct polar codes,'' \emph{IEEE Trans. Inf.
  Theory}, vol.~59, no.~10, pp. 6562--6582, Oct 2013.

\bibitem{ZamShaEre02}
R.~Zamir, S.~Shamai, and U.~Erez, ``Nested linear/lattice codes for structured
  multiterminal binning,'' \emph{IEEE Trans. Inf. Theory}, vol.~48, no.~6, pp.
  1250--1276, Jun 2002.

\end{thebibliography}
\end{document}